\documentclass[aps,twocolumn,floatfix,superscriptaddress]{revtex4-1} 
\usepackage{graphicx}
\usepackage{xcolor}
\usepackage{subfigure}
\usepackage{mathtools}
\usepackage{physics,amsmath}
\usepackage{epsfig}
\usepackage{chemfig}
\usepackage[colorlinks= true,urlcolor=blue,linkcolor= blue,citecolor=blue,bookmarks=false,pdfstartview=]{hyperref}
\usepackage{siunitx}
\usepackage{bm} 
\usepackage{xr}
\usepackage{xcolor}
\usepackage{placeins}
\usepackage{braket}
\usepackage{epsfig}
\begin{document}

\title{Magnetic properties of a staggered $S=1$ chain with an alternating single-ion anisotropy direction}

\author{S. Vaidya}
\email{s.vaidya@warwick.ac.uk}
\affiliation{Department of Physics, University of Warwick, Gibbet Hill Road, Coventry, CV4 7AL, UK}
\affiliation{School of Physics \& Astronomy, University of Birmingham, Edgbaston, Birmingham, B15 2TT, UK}
\author{S. P. M. Curley}
\affiliation{Department of Physics, University of Warwick, Gibbet Hill Road, Coventry, CV4 7AL, UK}
\author{P. Manuel}
\author{J. Ross Stewart}
\author{M. Duc Le}
\affiliation{ISIS Pulsed Neutron Source, STFC Rutherford Appleton Laboratory, Didcot, Oxfordshire OX11 0QX, United Kingdom}
\author{C. Balz}
\affiliation{ISIS Pulsed Neutron Source, STFC Rutherford Appleton Laboratory, Didcot, Oxfordshire OX11 0QX, United Kingdom}
\affiliation{Neutron Scattering Division, Oak Ridge National Laboratory, Oak Ridge, Tennessee 37831, USA}
\author{T.~Shiroka}
\affiliation{Center for Neutron and Muon Sciences, Paul Scherrer Institut, Forschungsstrasse 111, 5232 Villigen PSI, Switzerland}
\affiliation{Laboratorium für Festkörperphysik, Otto-Stern-Weg 1, ETH Zürich, CH-8093 Zurich, Switzerland}
\author{S. J. Blundell}
\affiliation{Department of Physics, Clarendon Laboratory, University of Oxford, Parks Road, Oxford, OX1 3PU, United Kingdom}
\author{K. A. Wheeler}
\affiliation{Department of Chemistry, Whitworth University, Spokane, Washington 99251, USA}
\author{I. Calderon-Lin}
\author{Z. E. Manson}
\author{J.~L.~Manson}\thanks{Deceased 7 June 2023.}
\affiliation{Department of Chemistry and Biochemistry, Eastern Washington University, Cheney, Washington 99004, USA}
\author{J. Singleton}
\affiliation{National High Magnetic Field Laboratory (NHMFL), Los Alamos National Laboratory, Los Alamos, NM, USA}
\author{T. Lancaster}
\affiliation{Department of Physics, Durham University, Durham DH1 3LE, United Kingdom}
\author{R. D. Johnson}
\affiliation{Department of Physics and Astronomy, University College London, Gower Street, London WC1E 6BT, United Kingdom}
\affiliation{London Centre for Nanotechnology, University College London, London WC1H 0AH, United Kingdom}
\author{P. A. Goddard}
\email{p.goddard@warwick.ac.uk}
\affiliation{Department of Physics, University of Warwick, Gibbet Hill Road, Coventry, CV4 7AL, UK}

\begin{abstract}
Materials composed of spin-1 antiferromagnetic (AFM) chains are known to adopt complex ground states which are sensitive to the single-ion-anisotropy (SIA) energy ($D$), and intrachain ($J_{0}$) and interchain ($J'_{i}$) exchange energy scales. While theoretical and experimental studies have extended this model to include various other energy scales, the effect of the lack of a common SIA axis is not well explored. Here we investigate the magnetic properties of Ni(pyrimidine)(H$_{2}$O)$_{2}$(NO$_{3}$)$_{2}$, a chain compound where the tilting of Ni octahedra leads to a 2-fold alternation of the easy-axis directions along the chain. Muon-spin relaxation measurements indicate a transition to long-range order at $T_{\text{N}}=2.3$\,K and the magnetic structure is initially determined to be antiferromagnetic and collinear using elastic neutron diffraction experiments. Inelastic neutron scattering measurements were used to find $J_{0} =  5.107(7)$\,K, $D = 2.79(1)$\,K, $J'_{2}=0.18(3)$\,K and a rhombic anisotropy energy $E=0.19(9)$\,K. Mean-field modelling reveals that the ground state structure hosts spin canting of $\phi\approx6.5^{\circ}$, which is not detectable above the noise floor of the elastic neutron diffraction data. Monte-Carlo simulation of the powder-averaged magnetization, $M(H)$, is then used to confirm these Hamiltonian parameters, while single-crystal $M(H)$ simulations provide insight into features observed in the data.

\end{abstract}

\maketitle

\section{Introduction}

Spin-1/2 Heisenberg antiferromagnetic (AFM) chains, in which the local spin environment periodically alternates in orientation, host magnetic properties that dramatically differ from their uniform counterparts. In systems such Cu(C$_{6}$D$_{5}$COO)$_{2}\cdot$3D$_{2}$O and Cu(pym)(H$_{2}$O)$_{2}$(NO$_{3}$)$_{2}$ (pym = pyrimidine C$_{4}$H$_{4}$N$_{2}$), two-fold staggered $g$ tensors and the Dzyalozhinskii-Moriya (DM) interaction result in a staggered internal field perpendicular to the applied field, leading to a field-induced spin-gap~\cite{Dender_Cu_benzo_1997, Feyerherm_stag_2000, Zvyagin_SG_2004, Huddart_spin_transport_2021}. This is in contrast to the gapless excitation spectra observed in conventional linear chains. The sine-Gordon model of quantum-field theory reproduces the staggered chain's excitation spectrum, which contains soliton and breather modes, plus a $\Delta\sim H^{2/3}$ field dependence of its energy gap~\cite{Oshikawa_SG_1997, Affleck_SG_1999}. 

Despite such unique phenomena displayed by non-linear $S=1/2$ chains, investigations into the effects of an alternating spin environment on AFM $S=1$ chains are notably lacking. In addition to the intra- and interchain energies $J_{0}$ and $J'$, the introduction of single-ion anisotropy (SIA) energy $D$ serves as an additional tuning parameter for $S=1$ systems. When considering a linear chain, the interplay between these parameters results in an already diverse set of possible magnetic ground states~\cite{Albuquerque_2009, Wierschem_2014}. In the ideal isotropic, one-dimensional limit, the Haldane gapped phase emerges, which hosts a topologically protected quantum disordered ground state~\cite{Haldane_1983}. With easy-plane anisotropy ($D>0$) as the $D/J_{0}$ ratio is increased, the Haldane phase is driven into a quantum paramagnetic phase. Increasing $J'$ drives the system into an $XY$-AFM ordered phase hosting a field-induced Bose-Einstein condensation of magnons~\cite{Zapf_BEC_2006, Zvyagin_2007}. In the easy-axis ($D<0$) case, Ising AFM order is induced~\cite{Wierschem_2014}. Further theoretical efforts have extended this model to explore the effects of rhombohedral anisotropy~\cite{Tzeng_2017}, non-Heisenberg exchange~\cite{Lange_XXZ_2018}, biquadratic exchange~\cite{Chiara_2011} and alternating exchange bonds~\cite{Suzuki_2018}. 
\begin{figure}
    \centering
    \includegraphics[width= \linewidth]{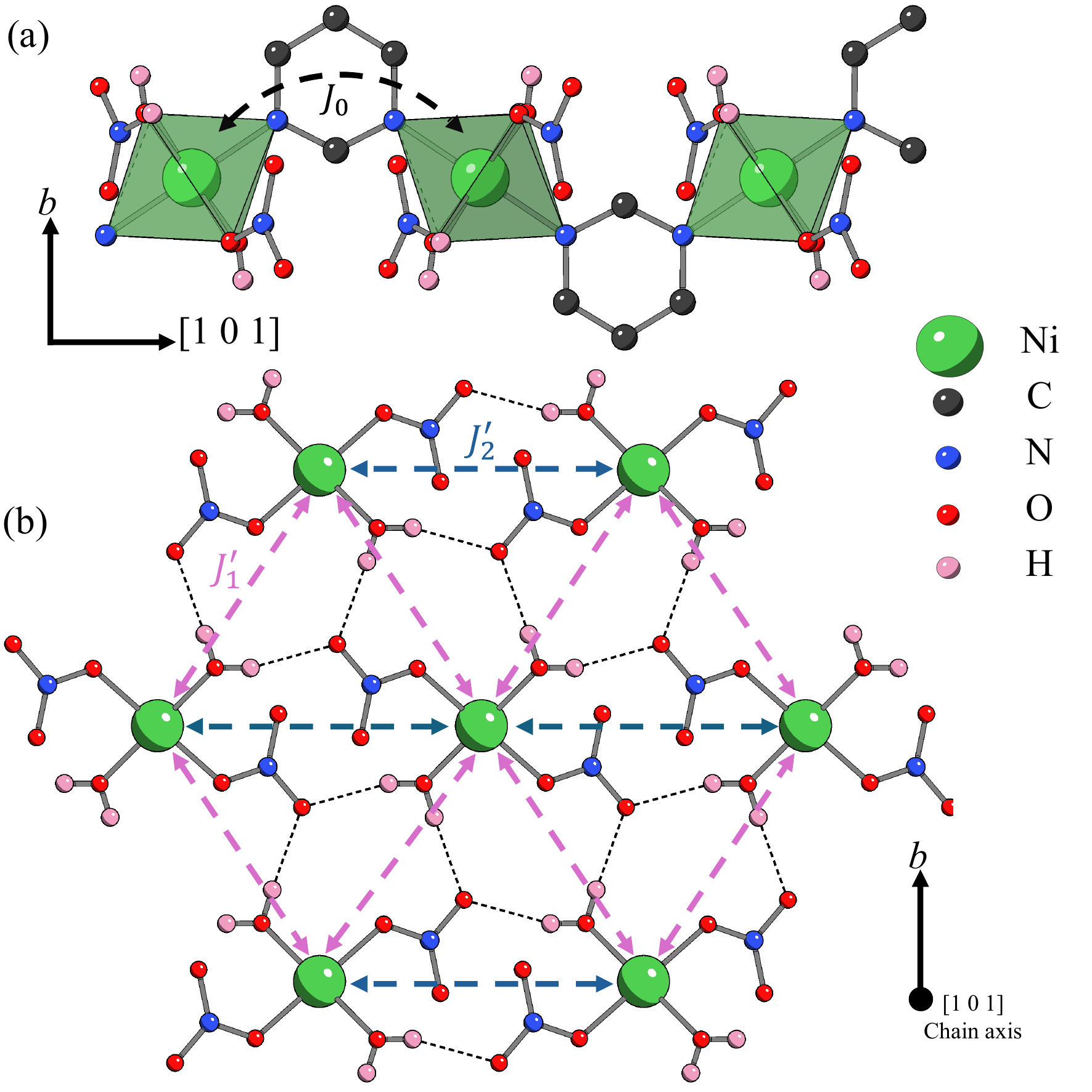}
    \caption{(a) The chain structure of Ni(pym)(H$_{2}$O)$_{2}$(NO$_{3}$)$_{2}$ (pym\,=\,pyrimidine) determined using single-crystal x-ray diffraction at $T=300$\,K. For clarity, $(\bar{1}\,0\,1)$ $hkl$-plane with only a single chain running along the $[1\,0\,1]$ direction is depicted. (b) View along the $[1\,0\,1]$ axis showing the hydrogen bond network connecting Ni(II) ions in adjacent chains. The purple and blue dashed arrows highlight the two types of inequivalent interchain exchange bonds $J'_{1}$ and $J'_{2}$. Each ion has six interchain neighbours.}
    \label{fig: Stag_struct}
\end{figure}

Here we turn our attention to Ni(pym)(H$_{2}$O)$_{2}$(NO$_{3}$)$_{2}$, an $S=1$ analogue of the $S=1/2$ staggered chain; in addition to the aforementioned staggered $g$ tensor, the alternating orientation of local octahedra now presents the possibility of a periodically alternating SIA axis. The alternating SIA, which lacks a global axis in these systems, is expected to act in competition with the exchange interaction. While a staggered SIA axis has attracted attention in the study of Glauber dynamics~\cite{SUN_2010-SCM, Zhang_Glauber_2013} and domain-wall dynamics~\cite{Pianet_2017_DW} in single-chain magnets, the ground state of quantum spin-chains with a staggered SIA axis does not appear to have been well explored theoretically or experimentally. In [Ni($\mu$-N$_{3}$(bmdt)(N$_{2}$]$_{n}$(DMF)$_{n}$ (bmdt =N,N'-bis(4-methoxylbenzyl)-diethylenetriamine, DMF
=N,N-dimethylformamide), weak ferromagnetism and unusual spin dynamics were reported. However, neutron scattering data were not available to solve the magnetic structure or determine the spin Hamiltonian~\cite{liu_2006_Ni-stag}. In three-dimensional and quasi-two-dimensional systems, an alternating easy-axis SIA direction has been found to induce large canting angles and drive weak ferromagnetism, which is normally caused by the DM interaction alone~\cite{Feyerherm_FeCl2_2004, XCl2L_Jem_2024}. In contrast, in the alternating easy-plane case, collinear order along a pseudo-easy-axis created by the intersection of the two local easy-planes is found~\cite{XCl2L_Jem_2024, Vaidya_2024_pseudo}. A canted low-temperature magnetic structure might be expected to occur in our Ni staggered chain, with the mean-field canting angle dictated by $D/J_{0}$. However, the low-temperature magnetic properties here are likely to be influenced by quantum fluctuations due to the reduced dimensionality and low spin quantum number. 

While these systems pose an interesting physical problem, the scarcity of suitably large single crystals and the comparable sizes of $D$ and $J_{0}$, are known, from the study of linear $S=1$ chains, to complicate the full characterisation of the Hamiltonian parameters~\cite{Brambleby_2017,blackmore_2019}. To overcome this, we follow a similar experimental protocol to that highlighted in Ref~\cite{Brambleby_2017}. We first use X-ray diffraction on small single crystals to determine the crystal structure and infer a magnetic Hamiltonian of the system. Magnetometry measurements on polycrystalline samples are used to confirm the quasi-low-dimensional nature of the interactions and easy-axis anisotropy. Muon-spin relaxation measurements find that the staggered chain undergoes long-range order below $T_{\rm N} = 2.2$\,K. While powder neutron diffraction reveals a co-linear magnetic ordering, the noise floor of the data leaves room for a large canting angle of up to $25^{\circ}$. Based on the constraints set on the Hamiltonian parameters by the magnetic structure, inelastic neutron scattering (INS) experiments were used to quantify these parameters which are then confirmed using Monte-Carlo (MC) simulation of the powder-averaged magnetization. 

\section{Results and Discussion}

\subsection{X-ray diffraction}
Single crystal X-ray diffraction was used to determine the crystal structure of the staggered chain, Ni(pym)(H$_{2}$O)$_{2}$(NO$_{3}$)$_{2}$, at $300$\,K with the structure shown in Fig.~\ref{fig: Stag_struct}. To minimize absorption corrections, small (sub mm) single crystals are preferred for these experiments. For our measurements we used a sample with dimensions $0.23\cross0.19\cross0.05$\,mm$^{3}$. Further details of the synthesis methods, single crystal XRD and structural refinement are provided in the supplementary material~\cite{supplementary}. Samples crystallise in the monoclinic space-group $C2/c$, with lattice parameters, $a = 12.7376(3)$\,$\si{\angstrom}$, $b = 11.4975(3)$\,$\si{\angstrom}$, $c = 7.3884(2)$\,$\si{\angstrom}$ and $\beta = 115.535(1)^{\circ}$. There are four Ni(II) ions residing at $[1/4\,1/4\,1/2]$, $[1/4\,3/4\,0]$, $[3/4\,3/4\,1/2]$ and $[3/4\,1/4\,0]$ positions in the unit cell. The octahedra of each Ni(II) ion include equatorial coordination to four O donor atoms from the H$_{2}$O and NO$_{3}$ ligands at distances $2.065$\,$\si{\angstrom}$ and $2.085$\,$\si{\angstrom}$, respectively, and axial coordination to N atoms from the pym ligands that bridge two Ni(II) ions related by a translation of $[1/2\,0\,1/2]$. Consequently, chains of Ni-pym-Ni lie along the crystallographic $[1\,0\,1]$ direction. However, due to the position of the N atoms in the aromatic pym ring, the octahedra of each Ni(II) ion are tilted out of the chain axis towards the $b$-axis by an angle $\alpha = 33.88(4)^{\circ}$, leading to a two-fold staggering of the Ni(II) octahedra as shown in Fig.~\ref{fig: Stag_struct}(a). From the structure, it is evident that any SIA present will result in the local N-Ni-N axis defining either the easy-axes or the normal to the easy-planes. Furthermore, the slight difference in the coordinate bonds to the two different ligand species  (H$_{2}$O and NO$_{3}$)  in the equatorial plane suggests the possibility of a small rhombohedral anisotropy with energy $E$. 

The hydrogen-bond networks illustrated in Figure~\ref{fig: Stag_struct}(b) stabilize the interchain structure. Each magnetic ion has six neighbours in adjacent chains; two neighbours residing at a distance of $7.388$\,$\si{\angstrom}$ along the $c$-axis, and four neighbours with distance $6.833$\,$\si{\angstrom}$ along the unit cell diagonals. Consequently, two possible interchain exchange interactions exist, $J'_{1}$ along the $[0,1,1]$ and $[0,\bar{1},1]$ directions and $J'_{2}$ along $c$-axis. These exchange pathways, depicted as purple and blue dashed arrows respectively in Fig.~\ref{fig: Stag_struct}(b), constitute a hexagonal lattice in the interchain directions. If $J'_{2}>0$ (AFM), this triangular arrangement of the interchain spins will result in competition between the $J'_{1}$ and $J'_{2}$ interactions.

The Hamiltonian of the staggered chain can be written as
\begin{equation}
    \label{eq: Hamitonian}   
\begin{split}
\mathcal{H}=&J_{0}\sum_{i}\hat{\mathbf{S}}_{i}\cdot\hat{\mathbf{S}}_{i+1}
     +\sum_{\left \langle i,j \right \rangle _{\perp}}J_{ij}'\hat{\mathbf{S}}_{i}\cdot\hat{\mathbf{S}}_{j}
     \\&+\sum_{i}\hat{\mathbf{S}}_{i}\cdot K_{i}\cdot\hat{\mathbf{S}}_{i}
    +\sum_{i}\mu_{\text{B}}\mu_{0}g\mathbf{H}\cdot\hat{\mathbf{S}}_{i},
\end{split}
\end{equation}
where $\hat{\mathbf{S}}_{i}$ is the spin of ion $i$. Here, the first sum is over the nearest neighbours along the Ni-pym chain with interaction strength $J_{0}$ and the second sum is over unique interchain exchange bonds $\left \langle i,j \right \rangle _{\perp}$. The third term describes the SIA with the local anisotropy tensors $K_{i}$ in the $xyz$ laboratory frame. In the local frame of Ni(II) ions, the anisotropy tensor $K^{loc}_{i} = \text{diag}\left[E,-E,D\right]$ and Euler rotations are used to transform $K^{loc}_{i}$ to $K_{i}$, as shown in the supplementary material~\cite{supplementary}. The final term is the Zeeman energy with an applied field $\mu_{0}\textbf{H}$ and an isotropic $g$-factor. In reality, we expect a small $g$-anisotropy of magnitude $\Delta g = g_{z} - g_{xy} = 2D/\lambda$, where $\lambda \sim500$\,K is a typical value of the spin-orbit coupling parameter for Ni(II) ions in octahedral environments~\cite{Boca_2004}. This would result in a staggered $g$-tensor and an internal staggered field with components perpendicular to the applied field. However, non-staggered Ni(II) systems with local environments similar to this material typically exhibit SIA energies on the order of $D\sim10$\,K~\cite{Manson_2020}, resulting in a very small $\Delta g \sim 0.04$, which is an order of magnitude smaller than seen in the Cu(II) staggered chains~\cite{Dender_Cu_benzo_1997, Feyerherm_stag_2000}.

C2/c is a centrosymmetric space group, with the Ni ions located on inversion centres, while the nearest-neighbour exchange bonds are not. This permits a DM interaction which changes sign from one bond to the next and has the form $(-1)^{i}\mathbf{D}_{m}\cdot\mathbf{S}_{i}\cross\mathbf{S}_{i+1}$. However, in molecule-based Ni(II) systems, where the spin-orbit coupling is relatively weak, the magnitude of the DM vector, $\mathbf{D}_{m}$, is expected to be small in comparison to the dominant interaction terms described in eq.~\ref{eq: Hamitonian}~\cite{Huang_2004, XCl2L_Jem_2024, Vaidya_2024_pseudo}.

\subsection{Magnetometry}
\begin{figure}
    \centering
    \includegraphics[width= \linewidth]{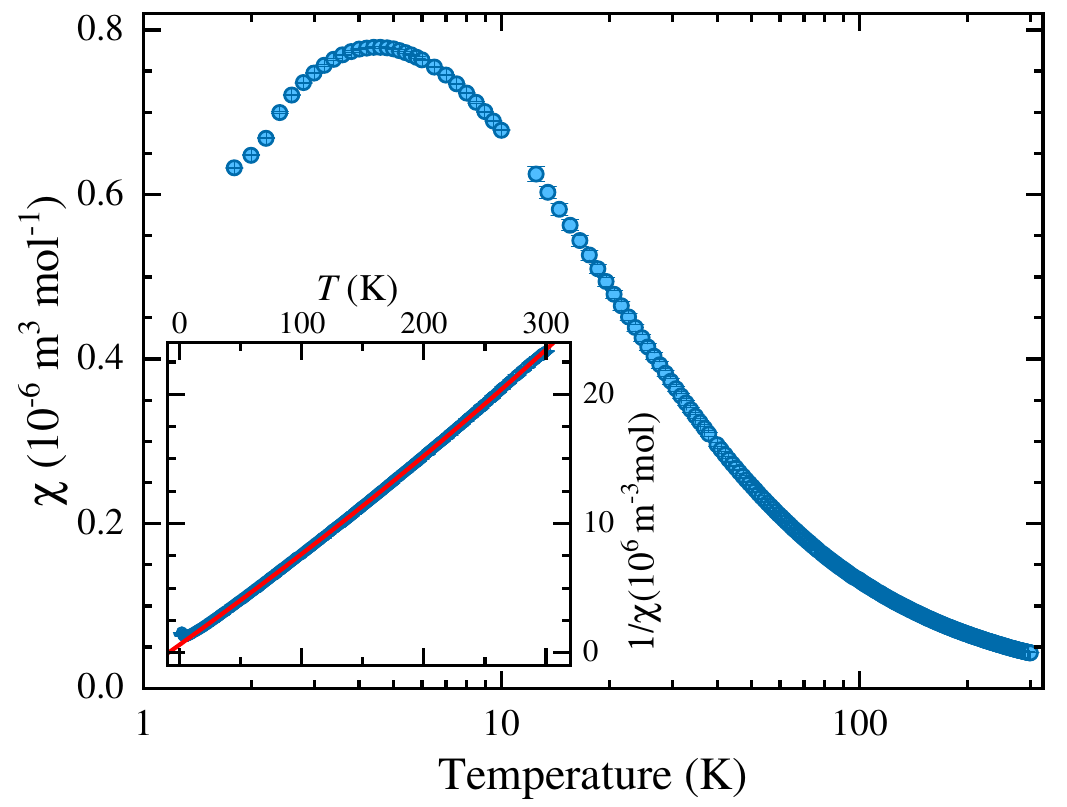}
    \caption{Temperature dependence of the zero-field-cooled (ZFC) magnetic susceptibility $\chi$ (blue circles) of the staggered $S=1$ chain Ni(pym)(H$_{2}$O)$_{2}$(NO$_{3}$)$_{2}$, measured in an applied field of $0.1$\,T. The field-cooled curve coincides with the ZFC. Inset: Data plotted as the inverse magnetic susceptibility $\chi^{-1}$ with a Curie-Weiss fit (red line).}
    \label{fig: Stag_Chi}
\end{figure}
\begin{figure}
    \centering
    \includegraphics[width= \linewidth]{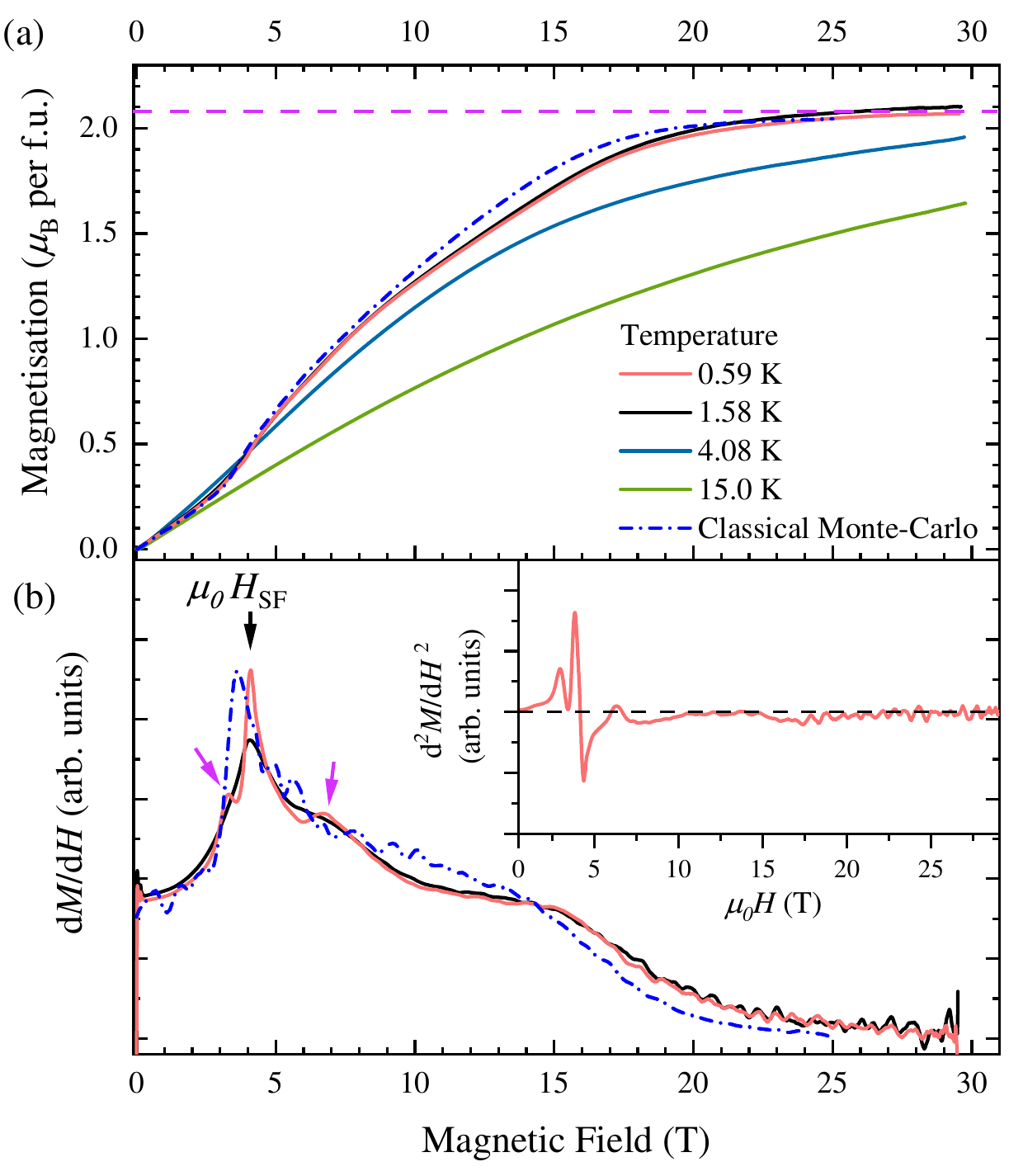}
    \caption{(a) Magnetization $M$ versus applied magnetic field $\mu_{0}H$ for powder samples of the staggered $S=1$ chain Ni(pym)(H$_{2}$O)$_{2}$(NO$_{3}$)$_{2}$ at various temperatures. The purple dashed line shows the $0.59$\,K magnetic saturation at $2.06$\,$\mu_{\text{B}}$\,per\,ion. (b) Differential susceptibility d$M$/d$H$ versus $\mu_{0}H$. The black arrow indicates the spin-flop transition seen as a peak in d$M$/d$H$. The purple arrows highlight the additional peaks observed in the $0.59$\,K data. The blue dash-dot line in both shows the results of a classical Monte-Carlo simulation. Inset: d$^{2}M$/d$H^{2}$ at $0.59$\,K.}
    \label{fig: Stag_MH}
\end{figure}

Fig.~\ref{fig: Stag_Chi} presents the zero-field-cooled (ZFC) magnetic susceptibility $\chi(T)$ data for powder samples of the staggered chain, Ni(pym)(H$_{2}$O)$_{2}$(NO$_{3}$)$_{2}$. The field-cooled $\chi(T)$ was found to coincide with the ZFC curve. For $75\leq T\leq 300$\,K, $\chi(T)$ is well described by the Curie-Weiss (CW) law, $\chi(T) = C/(T-\theta_{\text{CW}}) + \chi_{0}$, where $\chi_{0}$ is a temperature independent term, the Curie constant $C = N_{\text{A}}\mu_{0}g^{2}\mu_{\text{B}}^{2}S(S+1)/3k_{\text{B}}$ and $\theta_{\text{CW}}$ is the Curie-Weiss temperature. Fitting to $\chi^{-1}$, as shown in the inset to Fig.~\ref{fig: Stag_Chi}, gives $g = 2.18(1)$, $\chi_{0}=-5.49(7) \times 10^{-9}$\,m$^{3}$mol$^{-1}$ and $\theta_{\text{CW}} = -9.3(2)$\,K. The $g$-factor is typical of Ni(II) spin-1 systems, whereas the negative $\theta_{\text{CW}}$ is indicative of AFM coupling between Ni(II) ions. On cooling below $75$\,K, the data departs from the CW behaviour and develops a broad hump centred at around $T_{\chi \rm max} = 4.5(1)$\,K which is characteristic of quasi-low-dimensional systems and is due to the buildup of short-range correlations along the chain.

Figure~\ref{fig: Stag_MH}(a) and (b), respectively, present the pulsed-field magnetization, $M(H)$ curve and the differential susceptibility d$M$/d$H$ of powder samples at various temperatures. At the lowest temperature of $0.59$\,K, a sharp peak in d$M$/d$H$, corresponding to an upturn in magnetization, is observed at $\mu_{0}H_{\text{SF}} = 4.1(1)$\,T. We ascribe this feature to a spin-flop transition commonly observed in systems with easy-axis anisotropy. The transition is no longer present for $T  \geq 4.08$\,K, indicating long-range magnetic order is absent at this temperature. Additional satellite peaks are observed at $3.3(1)$\,T and  $6.7(2)$\,T which are broadened and no longer visible as peaks at $1.58$\,K. Increasing the field further results in a concave rise up to the projected magnetization saturation value of $m = 2.06$\,$\mu_{\text{B}}$\,per\,ion, suggesting a low-temperature $g =2.06(1)$. The $M(H)$ features and the Monte-Carlo simulation are discussed in detail in Sec.~\ref{sec: MC_sim}.

\subsection{Muon-spin relaxation}
\begin{figure}[h]
    \centering
    \includegraphics[width=\linewidth]{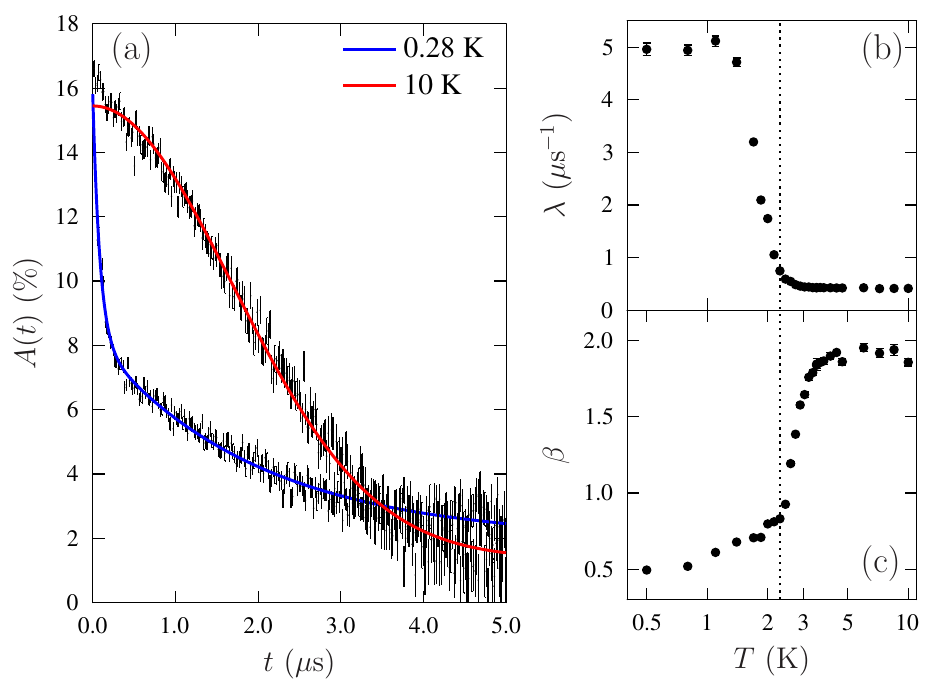}
    \caption{(a) Example ZF $\mu^{+}$SR spectra measured above and below the magnetic transition at $2.3$\,K. (b) Relaxation rate and (c) shape parameter extracted from fits to Eq~\ref{eq:fitting}. The estimated transition temperature is shown with a dotted line.}
    \label{fig: Stag_muon}
\end{figure}

In order to probe the magnetic transition, zero-field muon-spin relaxation (ZF $\mu^{+}$SR) measurements were performed on powder samples of Ni(pym)(H$_{2}$O)$_{2}$(NO$_{3}$)$_{2}$, down to $T=0.28$\,K. No oscillations are observed in the measured ZF spectra at any temperature; instead the spectra show rapid relaxation at low temperatures [Fig.~\ref{fig: Stag_muon}(a)]. At temperatures $T>5$\,K, the spectra show more gradual Gaussian relaxation, typically reflecting disordered electronic moments fluctuating too rapidly to relax the muon spins, leaving the muons to be depolarised by the quasistatic, disordered nuclear spins \cite{Blundell_muon-spec}. This is strongly suggestive of a magnetic ordering transition taking place between these limits in $T$. To capture the change in the spectra, data were fitted to the phenomenological function

\begin{equation}
A(t) = A_{1}{\rm e}^{-(\lambda t)^{\beta}} + A_{\mathrm{2}}. 
\label{eq:fitting}
\end{equation}
Here $\beta$ accounts for the evolving shape of the spectra and $\lambda$ parameterises the relaxation rate. The results of the fitting procedure are shown in Fig.~\ref{fig: Stag_muon}(b) and (c).

We see a sharply-defined change in the shape of the spectra and a rapid fall in $\lambda$ on warming from base temperature, consistent with the onset of long-range magnetic order at $T_{\mathrm{N}}=2.3$\,K. This also coincides with the end of the rapid fall in $\lambda$. (At low-$T$ in the presence of dynamic relaxation, $\lambda$ should be expected to vary with the square of the local fluctuating magnetic field, such that $\sqrt{\lambda}$ gives us a measure of the order parameter, assuming no change in fluctuation rate close to the transition \cite{Blundell_muon-spec}.)

We note that the presence of rapid relaxation in the muon spectra for $T<T_{\mathrm{N}}$, rather than oscillations, was also observed in related Ni(II)-based systems such as the series Ni$X_{2}$(pyz)$_{2}$~\cite{Liu_2016_correlations}, while oscillations are observed below $T_{\mathrm{N}}$ in other molecule-based systems, such as Ni(NCS)$_{2}$(thiourea)$_{2}$ \cite{Curley_2021}. The absence of oscillations here points to an increased level of magnetic disorder (compared to those cases where oscillations are observed), or to fast magnetic fluctuations on the muon timescale with a fluctuation rate $\nu>\gamma_{\mu}B_{\mathrm{int}}$, where $B_{\mathrm{int}}$ is the characteristic internal magnetic field at the muon site and $\gamma_{\mu}$ is the muon gyromagnetic ratio \cite{Blundell_muon-spec}. 

\subsection{Elastic neutron diffraction}

\begin{figure}
    \centering
    \includegraphics[width=\linewidth]{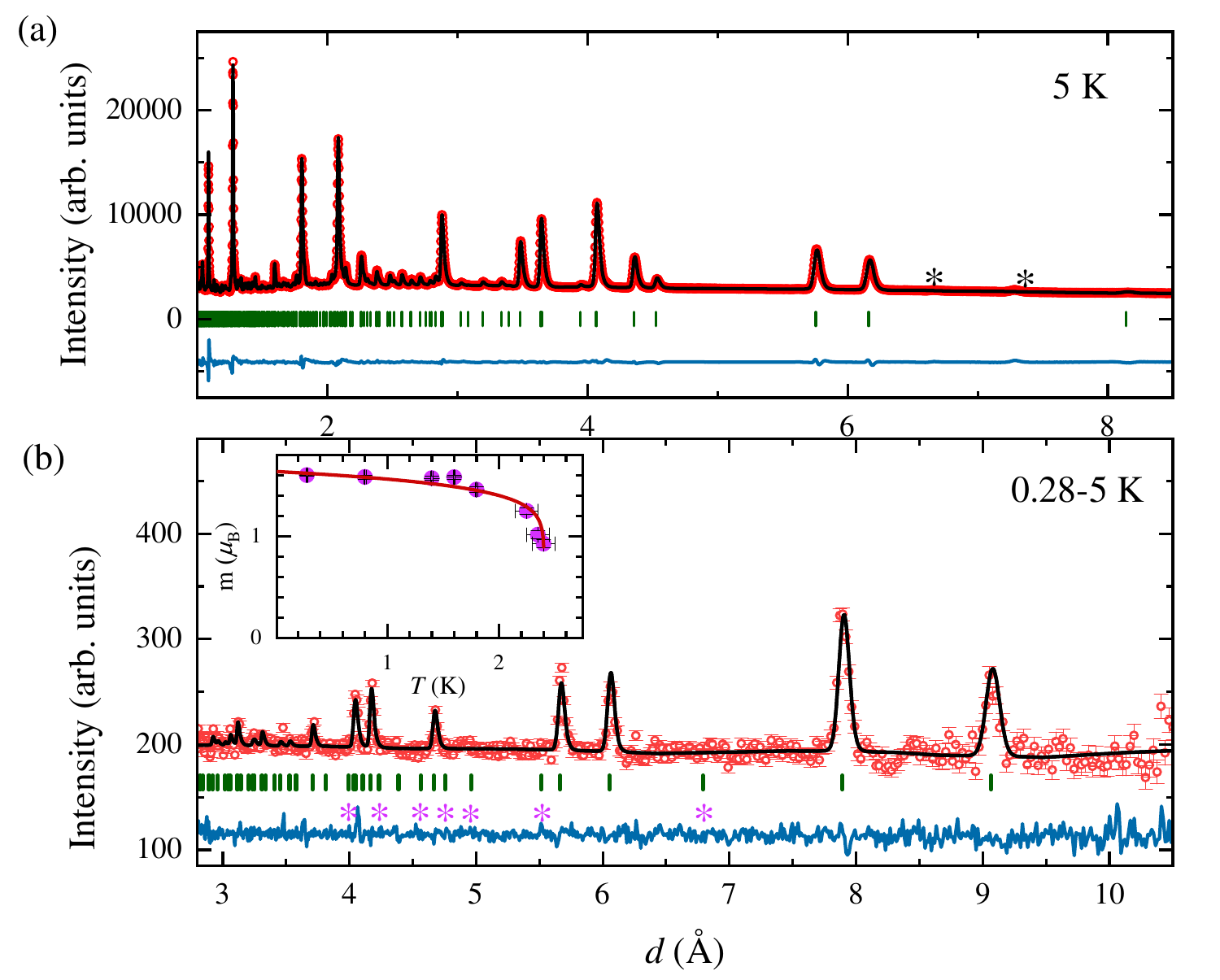}
\caption{Neutron powder diffraction data (red circles) of the staggered $S=1$ chain Ni(pym)(D$_{2}$O)$_{2}$(NO$_{3}$)$_{2}$.  Fits are shown as solid black lines, vertical tick marks show the Bragg position and solid blue lines indicate the difference curves. (a) Rietveld refinement of the $C2/c$ nuclear structure at $T=5$\,K in addition to a LeBail fit of a copper $Fm\bar{3}m$ phase. The black stars mark low-intensity impurity peaks. (b) The magnetic diffraction pattern, obtained by subtracting the nuclear and impurity peaks of the $T=5$\,K data from data collected at $T=0.28$\,K. Purple stars mark the ferromagnetic satellites of the $h+l = 2n+1$ Bragg positions which are discussed in the text but are absent in the data. Inset: $T$ dependence of the ordered Ni(II) magnetic moment.}
\label{fig: Stag_elastic}
\end{figure}

Elastic neutron diffraction measurements were performed on the WISH instrument at ISIS, the UK Neutron and Muon Source~\cite{Wish}. Data was collected on partially deuterated powder sample of Ni(pym)(D$_{2}$O)$_{2}$(NO$_{3}$)$_{2}$, where H$_{2}$O was substituted by D$_{2}$O. The magnetic properties were found to be very similar to the hydrogenated sample based on our $\chi(T)$ measurements. A quantitative Rietveld refinement of the crystal structure, conducted using FULLPROF~\cite{Fullprof}, is shown in Fig.~\ref{fig: Stag_elastic} (a) and reveals that the deuterated sample retains the $C2/c$ structure of the hydrogenated samples down to $0.28$\,K. The resulting lattice parameters, $a = 12.7441(3)$\,$\si{\angstrom}$, $b = 11.4933(4)$\,$\si{\angstrom}$, $c = 7.3871(2)$\,$\si{\angstrom}$ and $\beta = 115.488(3)^{\circ}$, are in close agreement with values obtained using X-ray measurements on the hydrogenated samples and the small differences are attributed to deuteration. At $2.08$\,$\si{\angstrom}$ and below, there are several additional high-intensity peaks modelled by a LeBail fit of a $Fm\bar{3}m$ copper structure. These peaks arise from the copper sample holder and a thin copper wire within the sample space required for cooling. Further details of the structural refinement are given in the supplementary material~\cite{supplementary}.

Subtracting the neutron diffraction data collected at $5$\,K from the data collected at $0.28$\,K reveals several additional peaks due to the long-range ordered magnetic structure [Fig.~\ref{fig: Stag_elastic}(b)]. Indexing these peaks reveals a commensurate magnetic propagation vector $\textbf{k}=(1/2,1/2,1/2)$. The observed magnetic peaks are satellites of the $h+k = 2n$ reflection allowed by $c$-centering. The projection of the centering vector, [1/2,1/2,0], on $\mathbf{k}$ yields a phase factor of $e^{i\pi}=-1$ on the moments of sites related by $c$-centering. Hence, the sites related by $c$-centering are aligned anti-parallel. Symmetry analysis using ISODISTORT~\cite{Isodistort, Isodistort_2006} reveals only one candidate symmetry (irrep $mL^{+}_{1}$) for the magnetic structure. Within this irreducible representation, the spins in the unit cell connected by pym (not related by $c$-centering) are symmetrically inequivalent magnetic sites and are not constrained by symmetry. The moments of these two spins can be decomposed into a linear combination of orthogonal ferromagnetic (FM), $\textbf{m}_{\text{FM}}$, and AFM, $\textbf{m}_{\text{AFM}}$, modes: 
\begin{equation}
    \textbf{m}_{1} = \textbf{m}_{\text{FM}} + \textbf{m}_{\text{AFM}}
\end{equation}
\begin{equation}
    \textbf{m}_{2} = \textbf{m}_{\text{FM}} - \textbf{m}_{\text{AFM}}.
\end{equation}

\begin{figure}
    \centering
    \includegraphics[width=\linewidth]{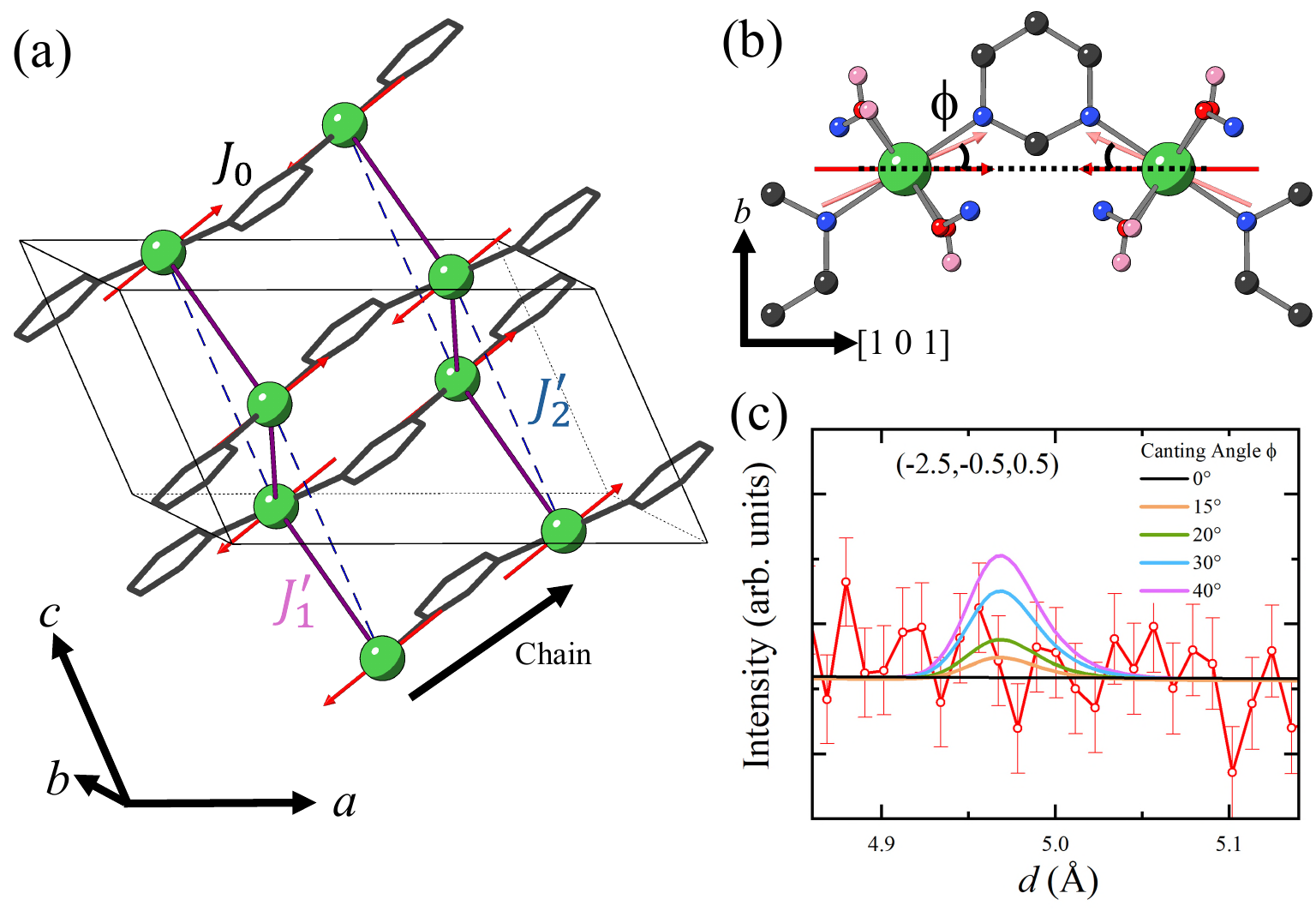}
\caption{ (a) The collinear magnetic structure of Ni(pym)(D$_{2}$O)$_{2}$(NO$_{3}$)$_{2}$, suggested by powder neutron diffraction measurements. The red arrows indicate the Ni(II) magnetic moment vectors. lines between ions highlights the three exchange interactions: $J_{0}$ through the pym ligand, $J'_{1}$ (solid purple lines) and $J'_{2}$ (blue dashed lines). The solid black line shows the monoclinic structural unit cell boundary. (b) View of the $(\bar{1}\,0\,1)$ $hkl$-plane. The pink arrows show the spin-canting at angle $\phi$, expected to be induced by a staggered easy-axis anisotropy direction. (c) Simulation of the $(-5/2, -1/2, 1/2)$ neutron diffraction peak for various canting angles with the observed data in red.}
\label{fig: Stag_mag_struct}
\end{figure}

As shown by the magnetic scattering intensity calculations in the supplementary material, $\textbf{m}_{\text{FM}}$ and $\textbf{m}_{\text{AFM}}$ can only contribute to the satellites of $h+l = 2n+1$ and $h+l = 2n$ peaks respectively~\cite{supplementary}. Only the AFM peaks are clearly present in the data of Fig.~\ref{fig: Stag_elastic}(b), suggesting a solely AFM coupling between spins connected by pym and absence of spin-canting arising from the FM mode. However, simulations of the most intense $h+l = 2n+1$ satellite peak, $(-5/2, -1/2, 1/2)$, in Fig.~\ref{fig: Stag_mag_struct}(c) show that for canting angles $\phi \leq 25 ^{\circ}$, these peaks would not be visible above the noise floor of the data. The magnetic refinement was carried out with a  fixed $\phi=0$.

The refined magnetic structure is presented in Fig.~\ref{fig: Stag_mag_struct}(a) and (b). Spins lie collinearly in the $ac$ plane at an angle, $\theta = 48(1)^{\circ}$ away from the $a$-axis. This structure indicates easy-axis anisotropy, where $\theta$ is expected to be determined by the projection of the local Ni-pym-Ni easy SIA axis onto the $ac$ plane, the rhombohedral anisotropy and a small allowed staggered DM interaction. By contrast, if there was a dominant staggered easy-plane anisotropy, then spins would be expected to align perpendicular to the chain and the $b$-axis, along a pseudo-easy-axis defined by the intersection of the two local easy-planes~\cite{XCl2L_Jem_2024, Vaidya_2024_pseudo}. 

If spin canting induced by a staggered easy-axis were present, the spins would rotate away from the $ac$ plane, towards the easy-axis (local Ni-N axis) at an angle of $\phi$ as depicted in Fig.~\ref{fig: Stag_mag_struct}(b). Considering a minimal mean-field model containing only $J_{0}$ and $D$ energy scales, the derivation presented in the supplementary material~\cite{supplementary} yields the expression:
\begin{equation}
    \frac{|D|}{J_{0}} = \frac{\sin(2\phi)}{\sin(\phi-\alpha)\cos(\phi-\alpha)}.
    \label{eq; DJ_ratio}
\end{equation}
The $\phi \leq 25 ^{\circ}$ limit of the data and Eq.~\ref{eq; DJ_ratio} set an upper limit on the ratio $|D|/J_{0} \leq 4.7$. 

Constraints on $J'_{1}$ and $J'_{2}$ are determined by the observed interchain order. Spins connected by the $J'_{2}$ exchange bond along the $c$-axis exhibit AFM order, which requires that $J'_{2}>0$ and $|J'_{2}|>|J'_{1}|$ (i.e. $J'_{2}$ must be AFM and stronger of than $J'_{1}$ for AFM order along $c$). If $|J'_{2}|<|J'_{1}|$, FM order along $c$ is imposed by the stronger $J'_{1}$ interaction and if $J'_{2}<0$, there is no competition between $J'_{1}$ and $J'_{2}$ leading to FM order along $c$. In our system, the competition between $J'_{1}$ and $J'_{2}$, is related to the formation of two $\textbf{k}$-domains, $\textbf{k}=(1/2,1/2,1/2)$ and $(-1/2,1/2,1/2)$. Depending on the choice of the $\textbf{k}$-domains, we find that the spins linked by the $J'_{1}$ exchange bonds exhibit either AFM or FM order along $[0,1,1]$ and FM or AFM order along $[0,\bar{1},1]$.

The temperature dependence of the low-temperature zero-field ordered Ni(II) moment, $m$, is shown in the inset to Fig.~\ref{fig: Stag_elastic}(b). Below $T=1.6$\,K, $m$ reaches a constant value of $1.6(1)$\,$\mu_{\text{B}}$. The suppression of the moment size, from the classical expectation value of $gS \approx 2.06$\,$\mu_{\text{B}}$, as seen at high-field, is indicative of the effect of quantum fluctuations. Additionally, the sharp change in $m$ at $T_{\rm N}$, is as expected for low dimensional systems~\cite{Uemura_1994, Keren_1995}. A fit to a power-law of the form $m(T) = m(0)(1 - T/T_{\text{N}})^{\delta}$ yields $m(0) = 1.64(3)$\,$\mu_{\text{B}}$, $\delta = 0.08(2)$ and $T_{\text{N}} = 2.40(2)$\,K, in good agreement with the magnetic ordering temperature determined through $\mu$SR measurements. Sparseness of the data points around $T_{\text{c}}$, where the critical exponent $\delta$ is most sensitive, means that the critical exponent $\delta$ extracted from this fit should not be used to give a tight constraint on theoretical models.

\subsection{Inelastic neutron scattering}

\begin{figure*}
    \centering
    \includegraphics[width= \textwidth]{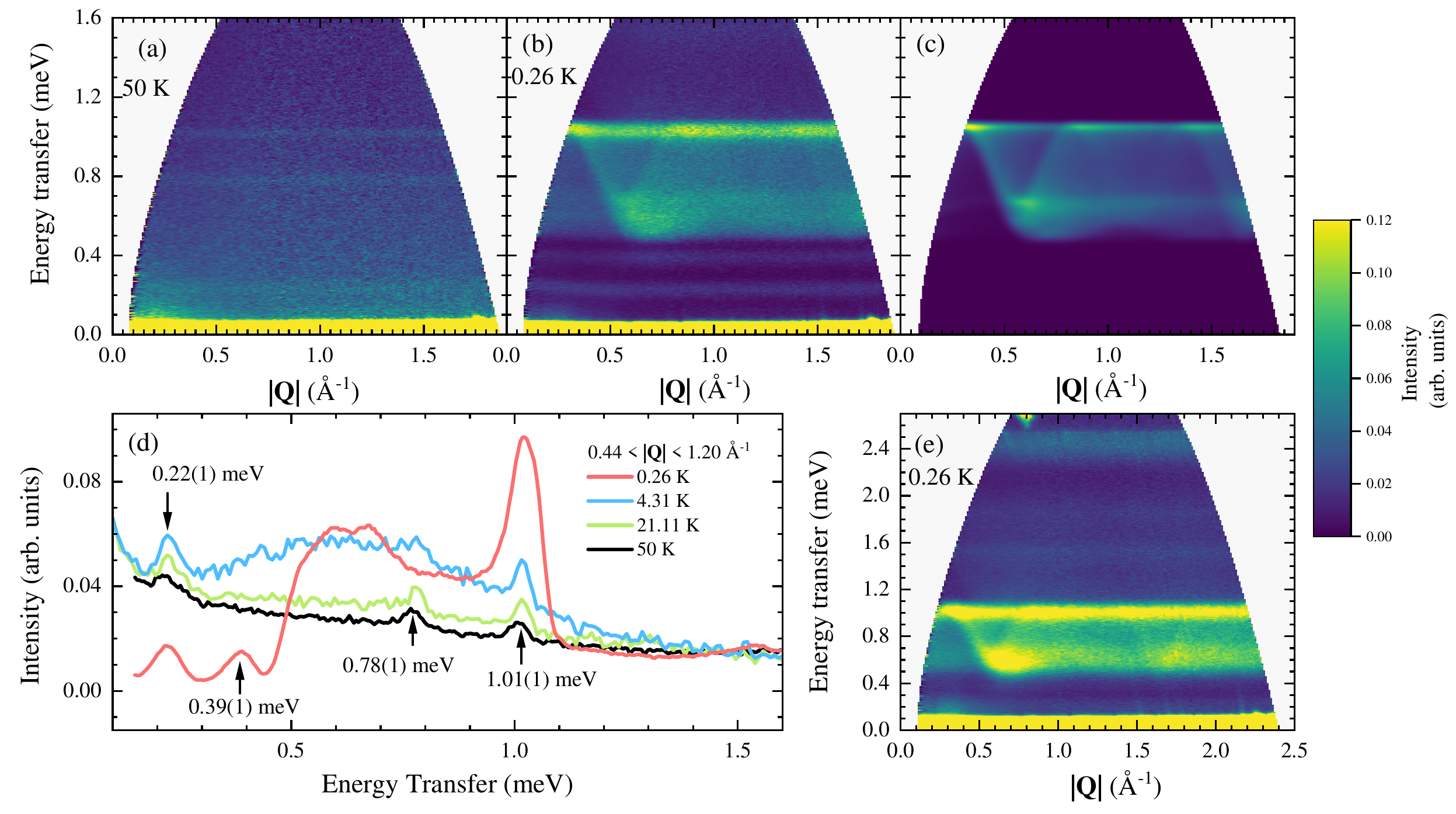}
\caption{Time-of-flight inelastic neutron scattering (INS) spectra measurement of powder Ni(pym)(D$_{2}$O)$_{2}$(NO$_{3}$)$_{2}$ with $\epsilon_{\text{i}} = 2.04$\,meV at (a) $T=50$\,K and (b) $T=0.26$\,K. (c) Powder average liner-spin-wave-theory simulation of the spin-wave spectra using Eq.~\ref{eq: Hamitonian} and the parameters: $J_{0}=5.107(7)$\,K, $J'_{1}= 0.00(5)$\,K, $J'_{2}=0.18(3)$\,K,  and single-ion-anisotropy with an easy axis component $D = -2.79(1)$\,K and a rhombohedral component $E=0.19(9)$\,K. (d) Temperature dependence of the neutron scattering intensity as a function of the energy transfer for Ni(pym)(H$_{2}$O)$_{2}$(NO$_{3}$)$_{2}$. The intensity here has been integrated over moment transfer $0.44\leq |\textbf{Q}|\leq 1.20$\,$\si{\angstrom}^{-1}$. (e) INS spectra collected at $T=0.25$\,K with $\epsilon_{\text{i}} = 3.36$\,meV.}
\label{fig: Stag_let_maps}
\end{figure*}

To quantify the sizes of the exchange interactions and anisotropy energies present in Ni(pym)(DO$_{2}$)$_{3}$(NO$_{3}$)$_{2}$, INS measurements were performed on the partially deuterated powder samples on the LET instrument at ISIS with incident energies $\epsilon_{i} = 2.04$ and $3.36$\,meV. LET is a multiplexing spectrometer which allows simultaneous measurements with different neutron incident energies. This allows us to survey a wide range of energy while maintaining high resolution at low energies. Further details are provided in Ref~\cite{LET, Let_data, Let_data_2}. Figure~\ref{fig: Stag_let_maps}(a) and (b) present the spectra collected at $T = 50$\,K, and $0.26$\,K respectively with incident energy $\epsilon_{\text{i}} = 2.04$\,meV. 
We note that additional measurements were performed at $T = 4.31$\,K and $21.11$\,K using high-resolution chopper settings, whilst the $T = 50$\,K and $0.26$\,K measurements used high flux chopper settings. This results in an intensity scaling factor of $\approx3.1$ which we obtained from calculations of the chopper opening times and verified by comparing the intensities of the nuclear (220) peak. In any case, the spectral features have widths $\approx0.1$\,meV which is considerably broader than the full-width at half-maximum (FWHM) resolution of either the high resolution ($0.035$\,meV) or high flux ($0.06$\,meV at the elastic line) modes, such that the datasets may be compared.

In the paramagnetic phase at $50$\,K, three dispersionless features are observed at neutron energy transfers of $\epsilon = 0.22(1)$, $0.78(1)$ and $1.01(1)$ meV. The momentum integrated energy cuts, at various temperatures, shown in Fig.~\ref{fig: Stag_let_maps}(d), reveal that the $0.78$\,meV excitation diminishes in intensity on cooling and is attributed to localised vibrational modes. In contrast, the $0.22$\,meV peak grows in intensity on cooling down to $0.25$\,K, implying it arises from a localised spin excitation from a ground state whose population increases on cooling. In this powder sample, this excitation could originate from the transition between the SIA split singlet $\ket{m_{s} = 0}$ and doublet $\ket{m_{s} = \pm 1}$ states of Ni(II) ions in an octahedral environment orphaned from the exchange network due to defects or chain ends. This would point to a SIA energy of $|D| = 2.6(1)$\,K. The $1.01$\,meV peak also grows in intensity on cooling and coincides with the large peak corresponding to the top of the spin-wave excitation band in the ordered phase.

Data collected in the ordered phase ($T=0.26$\,K) marks the emergence of dispersive spin-wave excitations along with a dispersionless in-gap excitation at $0.39(1)$\,meV [Fig.~\ref{fig: Stag_let_maps}(b)]. Additionally, in the $\epsilon_{i}=3.36$\,meV data, we observe low-intensity double-magnon scattering that extends to $1.8(2)$\,meV, approximately twice the value of the single-magnon excitation maximum, along with another dispersionless feature at $1.52(2)$\,meV. The spin-wave excitations were analysed using the Hamiltonian in Eq.~\ref{eq: Hamitonian} and linear spin wave theory (LSWT), as implemented in the SpinW~\cite{SpinW} program and taking the ground state from elastic neutron diffraction measurements. The SIA parameters used in SpinW simulations here were renormalised by a factor $\left [ 1 - 1/(2S) \right ]$ to account for the non-linear contributions to the SIA which are omitted in LSWT~\cite{Wheeler_2009_renorm, Dahlbom_2023}. 

Sharp dispersive features are observed in the data, extending from the energy gap of $0.46$\,meV and $0.69$\,$\si{\angstrom}^{-1}$ (magnetic Bragg position) to the top of the band at $1.02(1)$\,meV. Simulations using Eq.~\ref{eq: Hamitonian} indicate that the energy gap is predominantly governed by $D$ and increases with a larger $D$. Conversely, the bandwidth of the sharp dispersive feature narrows as $D$ increases and broadens with larger $J_{0}$. Reproducing the excitation gap, bandwidth and the sharp dispersion lines observed in the data allows us to accurately determine a unique set of values for $D$ and $J_{0}$, which are later refined against the data. 

Within the $0.46 \leq \epsilon \leq 0.76$\,meV region, there is an accumulation of spectral weight and a periodic modulation of the energy gap as a function of $|\mathbf{Q}|$. This suggests that intensity in this region arises from dispersive spin-wave branches that are broadened in $\epsilon$ and $|\mathbf{Q}|$ by powder-averaging. Indeed, we find that interchain spin-wave dispersions in the plane perpendicular to the chain are contained within this region. As such, increasing values of $J'_{2}$ increases the width of this band and the modulation of the gap as a function of $|\mathbf{Q}|$. Adding the $J'_{1}$ term results in the dispersion of the top of this band at $0.76$\,meV, which appears to be flat in our data, suggesting $J'_{1}$ is small. These features are well reproduced in our simulations, using estimated values of $J'_{2} \approx 0.18(3)$\,K and $J'_{1} \approx 0.00(5)$\,K. Simulations of the powder-average spin-wave spectra with different values of $J'_{1}$ and $J'_{2}$ are shown in the supplementary material~\cite{supplementary}. Inspection of the energy cut in Fig.~\ref{fig: Stag_let_maps}(d), reveals a broad feature in this region with two peaks at $0.60(2)$\,meV and $0.66(2)$\,meV, which was found to originate from to the rhombohedral anisotropy $E$.

\begin{figure*}
    \centering
    \includegraphics[width=0.9\linewidth]{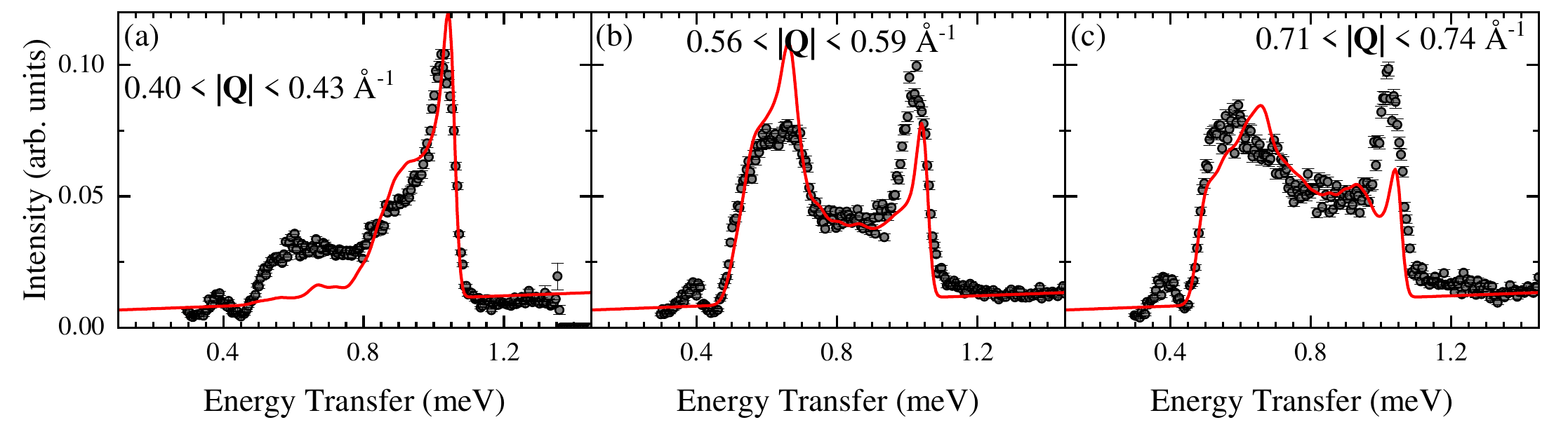}
\caption{ Inelastic neutron scattering intensity of the staggered $S=1$ chain Ni(pym)(D$_{2}$O)$_{2}$(NO$_{3}$)$_{2}$ as a function of the energy transfer, integrated over (a) $0.40\leq |\textbf{Q}|\leq 0.43$\,$\si{\angstrom}^{-1}$, (b) $0.56\leq |\textbf{Q}|\leq 0.59$\,$\si{\angstrom}^{-1}$ and (c) $0.71\leq |\textbf{Q}|\leq 0.74$\,$\si{\angstrom}^{-1}$. The data are shown as black circles and the red lines show the fits to the data discussed in the text.}
\label{fig: Stag_let_cuts_2}
\end{figure*}

The Levenberg-Marquardt (LM) algorithm was used to optimise $J_{0}$, $D$ and $E$ using seven $|\textbf{Q}|$ integrated energy cuts, three of which are shown in Fig.~\ref{fig: Stag_let_cuts_2}. Due to the differences in observed and simulated intensities discussed below, further optimisation of $J'_{1}$ and $J'_{2}$ was not possible and they were fixed to values estimated above. \footnote{When allowed to freely refine during the LM fitting, $J'_{1}$ and $J'_{2}$ approach $0$\,K and $|\mathbf{Q}|$-dependant modulation of the gap is no longer present in the simulations as shown in the supplementary material~\cite{supplementary}}. The resulting fitted parameters are $J_{0} =  5.107(7)$\,K ($0.4401(6)$\,meV), $D =-2.79(1)$\,K ($0.2412(1)$\,meV) and $E=0.19(9)$\,K ($0.016(8)$\,meV) and the simulation is depicted in Fig.~\ref{fig: Stag_let_maps}(c). The value of $D$ obtained through fitting the spin-wave spectra is in excellent agreement with the value suggested by the $0.22$\,meV dispersionless excitation and further supports the presence of localised excitations from orphaned Ni(II) ions as proposed earlier. Using Eq.~\ref{eq; DJ_ratio} and the fitted $|D|/J_{0} = 0.546(3)$, we estimate an canting angle $\phi \approx 6.5^{\circ}$. This canting is within the limit set by the noise floor of our elastic neutron diffraction data and is not expected to be discernable in those data. The AFM coupling causes spins in neighbouring chains to cant in opposite directions, cancelling the ferromagnetic component of a single chain. As a result, although spin-canting is present in these systems, a zero-field remanent magnetization is not expected, consistent with our powder $M(H)$ data. Additionally, the quasi-one-dimensional nature of our system, implied by our $\chi(T)$, $mu^{+}$SR and $T$-dependent neutron diffraction results, is confirmed by the ratio $|J'_{2}|/|J_{0}| \approx 0.04(2)$.

While the form of the observed spectra is captured very well by our model, there are differences in the observed and calculated intensities. This is most pronounced in the $0.46 \leq \epsilon \leq 0.76$\,meV region in the energy cut integrated over $0.40\leq |\textbf{Q}| \leq 0.43$\,$\si{\angstrom}^{-1}$ [Fig.~\ref{fig: Stag_let_cuts_2}(a)], which is dominated by the spin-wave dispersions in the plane perpendicular to the chain. Increased intensity observed in this region suggests a redistribution in the spectral weight from the spin-wave modes along the chain to the interchain modes. This may be attributed to the preferential orientation of the grains in the powder samples such that the interchain scattering plane is more exposed to the neutron beam.
Additionally, because LSWT does not account for impurity effects, the $0.22$\,meV feature is not reproduced in our simulation. We also note that the $0.39$\,meV in-gap modes were not reproduced in the SpinW simulations, even when higher-order interactions are included, and could hint at a localised excitation mode which LSWT does not capture. One possible explanation is that this feature could arise from excitations between energy levels within small orphaned dimers of Ni(II) ions~\cite{Lee_1996}. Other orphaned structures, such as trimers, may also account for the 1.01 meV excitation seen up to $50$\,K.

\subsection{Monte-Carlo simulation of $M(H)$} \label{sec: MC_sim}

\begin{figure}
    \centering
    \includegraphics[width=\linewidth]{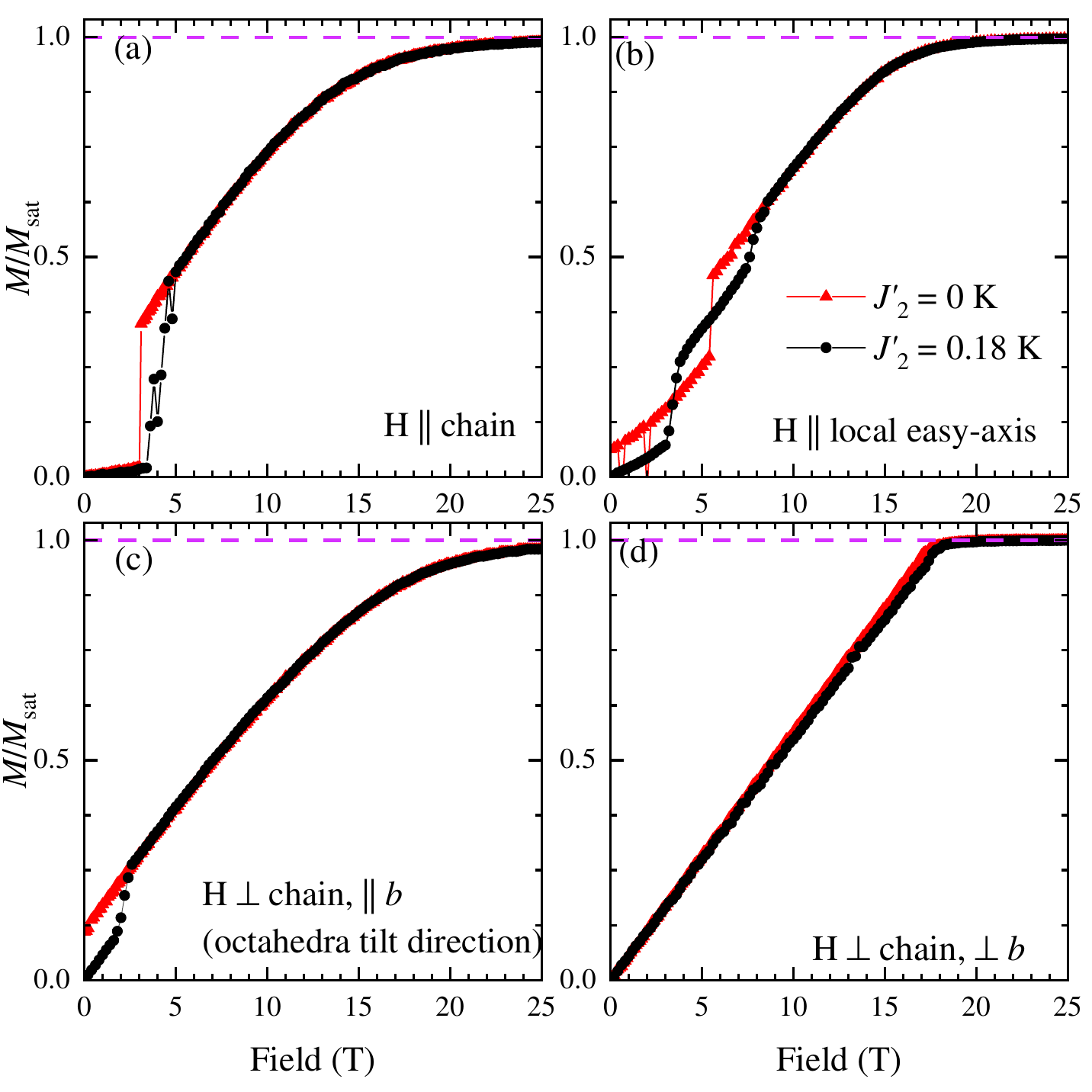}
\caption{Monte-Carlo simulation of the field-dependent magnetization of the staggered $S=1$ chain at various field orientations, using Hamiltonian parameters determined using inelastic neutron scattering. The field is applied (a) parallel to the chain, (b) parallel to a local easy-axis of a Ni(II) spin, (c) perpendicular to the chain and parallel to the $b$-axis (octahedra tilting direction), and (d) perpendicular to the chain and $b$ (perpendicular to the plane containing the easy-axes). The purple dashed line in all panels indicates magnetic saturation.}
\label{fig: Stag_MC}
\end{figure}

Monte-Carlo simulations of the powder-averaged longitudinal $M(H)$ were performed to confirm the Hamiltonian parameters determined by the INS experiments. At each field, the simulation aims to minimise the total energy of an 8-spin cluster, which is calculated using the Eq.~\ref{eq: Hamitonian} and an isotropic $g=2.06$ that was determined from the $M(H)$ data at $T=0.59$\,K. The resulting $M(H)$ and d$M$/d$H$ simulations are shown as blue dash-dot lines in Fig.~\ref{fig: Stag_MH}. The simulations provide an overall good agreement with the observed spin-flop field at $\mu_{0}H_{\rm SF} = 4.1(1)$\,T and the rounded approach to the saturation at high fields. 

Simulations of single-crystal $M(H)$ data with $J'_{2}=0$\,K and $J'_{2}=0.18$\,K ($J'_{1}=0$\,K in both cases) were performed to assess the effects of interchain interactions on $M(H)$. The $J'_{2}=0$\,K simulations consider only a single isolated spin chain, and as such show a zero-field remanent magnetization, which as explained above, disappears when multiple chains are considered. When the applied field is parallel to the chain [see Fig.~\ref{fig: Stag_MC}(a)], a single spin-flop transition occurs for both $J'_{2}=0$\,K and $J'_{2}=0.18$\,K. This transition occurs at $\mu_{0}H_{\rm SF}=4$\,T for $J'_{2}=0.18$\,K and corresponds to the main spin-flop feature found in the power average $M(H)$ data.

In Fig.~\ref{fig: Stag_MC}(b), where the applied field is parallel to one of the easy-axis directions, there is a single meta-magnetic transition for $J'_{2}=0$\,K and two separate transitions at $3.4$\,T and  $7.6$\,T when $J'_{2}=0.18$\,K. Similarly, in Fig.~\ref{fig: Stag_MC}(c), where the applied field is parallel to the $b$-axis, the octahedra tilting direction, a metamagnetic transition, associated with neighbouring chains flipping, is present at $2.2$\,T only when $J'_{2}=0.18$\,K.

These additional metamagnetic transitions, which appear in the simulations for certain field directions when interchain interactions are considered, may account for the $3.3(1)$\,T and  $6.7(2)$\,T, satellite features observed in the $0.59$\,K powder-averaged $dM/dH$ data. In a powder-averaged measurement, a single peak in $dM/dH$ is expected, as indicated by the powder-average MC simulations. Preferential orientation of the grains during measurements, may result in an increased contribution to $dM/dH$ from these directions and the observation of distinct satellite peaks in our data.

In anisotropic magnets, a symmetry-breaking phase transition is often expected to occur at saturation when a large field is applied parallel to the principle anisotropy axis. However, it has been shown recently that in systems with an alternating SIA axis, there are principle directions where the SIA can always save energy by canting spins away from the applied field, and magnetic saturation only occurs in the infinite field limit~\cite{Vaidya_2024_pseudo}. Similarly in Ni(pym)(H$_{2}$O)$_{2}$(NO$_{3}$)$_{2}$, a magnetic saturation phase transition only occurs when the applied field is simultaneously perpendicular to the chain and the $b$-axis [Fig.~\ref{fig: Stag_MC}(d)]. This is the direction about which the octahedra rotate and is the only direction where the applied field is perpendicular to both easy-axes. For all other field orientations, $M(H)$ asymptotically approaches the saturation value. In the powder-averaged $M(H)$ simulation and data, this is captured by the rounded approach to saturation which shows no features in $dM/dH$ indicative of a phase transition.

\section{Conclusions}

We have explored the magnetic properties of a quasi-one-dimensional $S=1$ material Ni(pym)(H$_{2}$O)$_{2}$(NO$_{3}$)$_{2}$. The octahedra of neighbouring Ni(II) ions, connected by pym ligands, alternate in orientation leading to a staggered Ni-pym chain. The alternating SIA direction is expected to follow the Ni(II) octahedra. Muon-spin relaxation data show signatures of long-range order below $T_{\text{N}} = 2.3$\,K, which is stabilised by non-zero interchain interactions. Powder elastic neutron diffraction measurements reveal AFM order and are consistent with a staggered easy-axis anisotropy. In systems with an alternating easy-axis direction, spin canting towards the local easy-axis direction is expected~\cite{Feyerherm_FeCl2_2004, XCl2L_Jem_2024}. The noise floor of the neutron diffraction data leaves room for spin-canting of up to $25^{\circ}$. Powder inelastic neutron scattering experiments were performed to quantify the Hamiltonian. Its parameters, $J_{0} =  5.107(7)$\,K, $J'_{1} \approx 0.00(5)$, $J'_{2} \approx 0.18(3)$\,K, $D =-2.79(1)$\,K and $E=0.19(9)$\,K were then determined using INS measurements on powder samples and for the experimentally determined ratio $|D|/J_{0} = 0.55(1)$, mean-field calculations show that $\phi\approx6^{\circ}$. While our LSWT simulations model the observed dispersive spin-wave excitations well, there are certain features which are not captured. We suggest that some of these features arise from small clusters of Ni(II) spins, such as single ions, dimers and trimers which are orphaned from the chains. Additionally, it has been pointed out elsewhere that some details of $S = 1$ magnetic excitation spectra can only be accounted for generalizing spin–wave modeling to include SU(3) degrees of freedom~\cite{bai_2021}. This is a developing issue that requires further investigation.

Monte-Carlo simulations, using Eq.~\ref{eq: Hamitonian} and the Hamiltonian parameters determined using INS measurements, show that the features in the $M(H)$ are largely accounted for by a classical model and an isotropic $g$-factor. This is in contrast to the $S=1/2$ staggered chain, where the $g$-anisotropy and the DM interactions have a dramatic effect, mapping the Hamiltonian on to the sine-Gordon model of quantum field theory and giving rise to soliton-like excitation modes~\cite{Dender_Cu_benzo_1997, Feyerherm_stag_2000, Zvyagin_SG_2004, Huddart_spin_transport_2021, Oshikawa_SG_1997, Affleck_SG_1999}. Recently, experiments on a chiral $S=1/2$ system, [Cu(pym)(H$_{2}$O)$_{4}$]SiF$_{6}\cdot$H$_{2}$O, hosting a four-fold periodic variation of the spin environments showed behaviour which was not governed by the sine-Gordon model~\cite{Liu_chiral_2019}. Therefore, exploring a similar extension from the $S=1$ staggered chain to an $S=1$ chiral chain could prove interesting. 

For a linear $S=1$ AFM chain with $|D|/J_{0} = 0.546(3)$ and $|J'_{ij}|/|J_{0}| \approx 0.04(2)$, the Haldane phase is already expected to be quenched by $D$ and $J'$ into an AFM Ising order~\cite{Sengupta_2014}. Therefore, to study the robustness of the topological phase in the presence of alternating octahedra and single-ion anisotropy direction, further theoretical studies and the development of real materials close to the quantum critical point are necessary.

\begin{acknowledgments}
We are indebted to the late Jamie Manson for instigating this work, for his role in designing and growing the samples and for many other invaluable contributions. We thank T. Orton and P. Ruddy for their technical assistance. We also thank R. Coldea, D. M. Pajerowski, C. Stock and J. A. Villa for valuable discussions and B. M. Huddart for his assistance with the muon measurements. SV thanks the UK Engineering and Physical Sciences Research Council (EPSRC) for supporting his studentship. This project has received funding from the European Research Council (ERC) under the European Union’s Horizon 2020 research and innovation programme (Grant Agreement No. 681260) and EPSRC (Grant No. EP/N024028/1). A portion of this work was performed at the National High Magnetic Field Laboratory, which is supported by National Science Foundation Cooperative Agreements Nos. DMR-1644779 and DMR-2128556, the US Department of Energy (DoE) and the State of Florida. JS acknowledges support from the DoE BES FWP “Science of 100 T". Part of this work was carried out at the Swiss Muon Source, Paul Scherrer Institut and we are grateful for the provision of beamtime. Data presented in this paper will be made available at XXX and the INS data can be found in Refs.~\cite{Let_data, Let_data_2}. For the purpose of open access, the author has applied a Creative Commons Attribution (CC-BY) licence to any Author Accepted Manuscript version arising from this submission.
\end{acknowledgments}

\bibliography{main.bib}
\end{document}


\title{Supplementary Material accompanying Magnetic properties of a staggered $S=1$ chain with an alternating single-ion anisotropy direction}

\author{S. Vaidya}
\email{s.vaidya@warwick.ac.uk}
\affiliation{Department of Physics, University of Warwick, Gibbet Hill Road, Coventry, CV4 7AL, UK}
\affiliation{School of Physics \& Astronomy, University of Birmingham, Edgbaston, Birmingham, B15 2TT, UK}
\author{S. P. M. Curley}
\affiliation{Department of Physics, University of Warwick, Gibbet Hill Road, Coventry, CV4 7AL, UK}
\author{P. Manuel}
\author{J. Ross Stewart}
\author{M. Duc Le}
\affiliation{ISIS Pulsed Neutron Source, STFC Rutherford Appleton Laboratory, Didcot, Oxfordshire OX11 0QX, United Kingdom}
\author{C. Balz}
\affiliation{ISIS Pulsed Neutron Source, STFC Rutherford Appleton Laboratory, Didcot, Oxfordshire OX11 0QX, United Kingdom}
\affiliation{Neutron Scattering Division, Oak Ridge National Laboratory, Oak Ridge, Tennessee 37831, USA}
\author{T.~Shiroka}
\affiliation{Center for Neutron and Muon Sciences, Paul Scherrer Institut, Forschungsstrasse 111, 5232 Villigen PSI, Switzerland}
\affiliation{Laboratorium für Festkörperphysik, Otto-Stern-Weg 1, ETH Zürich, CH-8093 Zurich, Switzerland}
\author{S. J. Blundell}
\affiliation{Department of Physics, Clarendon Laboratory, University of Oxford, Parks Road, Oxford, OX1 3PU, United Kingdom}
\author{K. A. Wheeler}
\affiliation{Department of Chemistry, Whitworth University, Spokane, Washington 99251, USA}
\author{I. Calderon-Lin}
\author{Z. E. Manson}
\author{J.~L.~Manson}\thanks{Deceased 7 June 2023.}
\affiliation{Department of Chemistry and Biochemistry, Eastern Washington University, Cheney, Washington 99004, USA}
\author{J. Singleton}
\affiliation{National High Magnetic Field Laboratory (NHMFL), Los Alamos National Laboratory, Los Alamos, NM, USA}
\author{T. Lancaster}
\affiliation{Department of Physics, Durham University, Durham DH1 3LE, United Kingdom}
\author{R. D. Johnson}
\affiliation{Department of Physics and Astronomy, University College London, Gower Street, London WC1E 6BT, United Kingdom}
\affiliation{London Centre for Nanotechnology, University College London, London WC1H 0AH, United Kingdom}
\author{P. A. Goddard}
\email{p.goddard@warwick.ac.uk}
\affiliation{Department of Physics, University of Warwick, Gibbet Hill Road, Coventry, CV4 7AL, UK}

\maketitle
\tableofcontents
\section{Further Experimental Details}
\subsection{Synthesis}

All chemical reagents were obtained from commercial sources and used as received. All reactions were performed in 25 ml glass beakers. For Ni(pym)(H$_{2}$O)$_{2}$(NO$_{3}$)$_{2}$, an aqueous solution of (NO$_{3}$)$_{2}$  ($0.300$\,g, $1.032$\,mmol) was slowly added to an aqueous solution of pyrimidine ($0.0826$\,g, $1.032$\, mmol) to afford a clear blue solution. Drops $0.1$\,M acetic acid was then added to the solution, the resulting solution remained a clear blue. Slow evaporation of the solution at room temperature yields small clear green crystallites. 

\subsection{Single-crystal X-ray diffraction}

Single-crystal x-ray diffraction measurements were conducted on Ni(pym)(H$_{2}$O)$_{2}$(NO$_{3}$)$_{2}$ crystals measuring $0.23 \cross 0.19 \cross 0.05$\,mm$^{3}$ at $T=100$\,K. Data were collected using a Bruker Venture D8 diffractometer equipped with a Bruker PHOTON II detector and Mo K$_{\alpha}$ ($0.71073$\,$\si{\angstrom}$) radiation generated by a microfocus sealed X-ray tube. Measured intensities were corrected for absorption using the Multi-Scan method (SADABS)~\cite{Krause_multi-scan}. The structure was solved using ShelXT~\cite{SheldrickT} and refined using ShelXL~\cite{SheldrickL} as implemented through Olex2~\cite{Dolomanov_olex}. C-H hydrogen atoms were restrained to calculated distances, while O-H hydrogen atoms were located in the Fourier difference maps. All non-H atomic displacements were treated anisotropically while those of hydrogen atoms were treated isotropically and riding on the donor atom. Table~\ref{tab: X-ray} provides full details of the structural refinement. The structure has been deposited and may be found at Deposition Number CCDC 2403842.

\begin{table}
\caption{\label{tab: X-ray} Single Crystal X-ray data and refinement details for Ni(pym)(H$_{2}$O)$_{2}$(NO$_{3}$)$_{2}$.}
\begin{tabular}{p{6cm}p{5cm}}
\hline\hline
\multicolumn{2}{c}{Crystal data}\\
\colrule
Empirical formula                &NiC$_{4}$H$_{8}$N$_{4}$O$_{8}$ \\
$T$\,(K)                    & $100$ \\
Crystal system              & Monoclinic \\
Space group                 & $C2/c$  \\
$a,b,c$\,$(\si{\angstrom})$     & $12.7376(3)$, $11.4975(4)$, $7.884(2)$\\
$\beta$                         & $115.535(1)^{\circ}$\\
$V\,(\si{\angstrom}^{3})$ (as used)	&$976.34(4)$\\
$Z$                         &4\\
Crystal size (mm$^{3}$) & $0.23 \cross 0.19 \cross 0.05$\\
\colrule
\multicolumn{2}{c}{Data collection}\\
\colrule
Instrument & Bruker Venture D8 diffractometer \\
Radiation & Mo K$_{\alpha}$ ($0.71073$\,$\si{\angstrom}$)\\
No. of measured reflections & 7520\\
No. of independent reflections  & 1001 \\
No. of measured reflections [$I\geq2\sigma(I)$]& 898\\
$R_{\text{int}}$                &$0.05$  \\
$h$-index range & $-15\leq h \leq 15$\\
$k$-index range & $-14\leq k \leq 14$\\
$l$-index range & $-9\leq l \leq 9$\\
Absorption Correction & Multi-Scan method (SADABS) \\
\colrule
\multicolumn{2}{c}{Refinement}\\
\colrule
\multirow{3}{*}{R indexes [$F^{2}\geq2\sigma(F^{2})$]}&R\,=\,$0.019$ \\                                                                            &$w$R\,=\,$0.0452$ \\
                                                      &$S$\,=\,$1.118$\\
Data/param./restr. &$1001$/$88$/$1$\\
$\Delta\rho_{\text{max}}$, $\Delta\rho_{\text{max}}$\,(e$\si{\angstrom}^{-3}$)&$0.3$, $-0.35$\\
\hline\hline
\end{tabular}
\end{table}

\subsection{Magnetometry}

\paragraph{SQUID} Measurements of $\chi(T)$ were performed using a Quantum Design MPMS XL SQUID magnetometer. Powder samples of Ni(pym)(H$_{2}$O)$_{2}$(NO$_{3}$)$_{2}$ were suspended in Vaseline to mitigate the movements of grains in the field. The Vaseline-sample mix is held within a gelatine capsule with a low magnetic background. The gelatine capsule is then placed in a plastic drinking straw for the measurement. Samples were cooled to $T=1.8$\,K in zero-field and zero-field-cooled (ZFC) measurements were performed on warming to $T=300$\,K with a constant applied field of $\mu_{0}H = 0.1$\,T. Field-cooled measurements were then performed on cooling back down to $T=1.8$\,K. $M(H)$ measurements at constant temperatures of $T=1.8$\,K, $4$\,K and $10$\,K were performed for the ZFC samples.

\paragraph{Pulsed-field} Isothermal pulsed-field magnetization measurements were performed at the National High Magnetic Field Laboratory in Los Alamos, USA. Fields of up to $65$\,T with a typical rise time of $\approx10$\,ms were used. Polycrystalline samples were packed into a PCTFE ampoule (inner diameter $1.0$\,mm) and sealed with vacuum grease to prevent sample movement. The ampoule can be moved in and out of a $1500$-turn, $1.5$\,mm bore, $1.5$\,mm-long compensated-coil susceptometer constructed from $50$-gauge high-purity copper wire. When the sample is in the coil, the voltage induced in the coil is proportional to the rate of change in $M$ over time (d$M$/d$t$). The signal is integrated and the background data, measured with an empty coil under the same conditions, is subtracted to obtain $M(H)$. The magnetic field value is measured using a coaxial $10$-turn coil and calibrated using observations of de Haas-van Alphen oscillations arising from the copper coils of the susceptometer. A $^{3}$He cryostat provides temperature control and is used to attain temperatures down to $500$\,mK. The $M(H)$ measured in pulsed-fields were normalised using $T=1.8$\,K and $4$\,K data measured in the SQUID magnetometer.

\subsection{Muon-spin relaxation}
Zero-field muon-spin relaxation measurements were made on powder samples of Ni(pym)(H$_{2}$O)$_{2}$(NO$_{3}$)$_{2}$ using the Dolly spectrometer at the Swiss Muon Source, Paul Scherrer Institut, Switzerland, using a $^{3}$He insert to attain temperatures down to $0.28$~K.

\subsection{Elastic neutron diffraction}

\begin{figure}
    \centering
    \includegraphics[width=\linewidth]{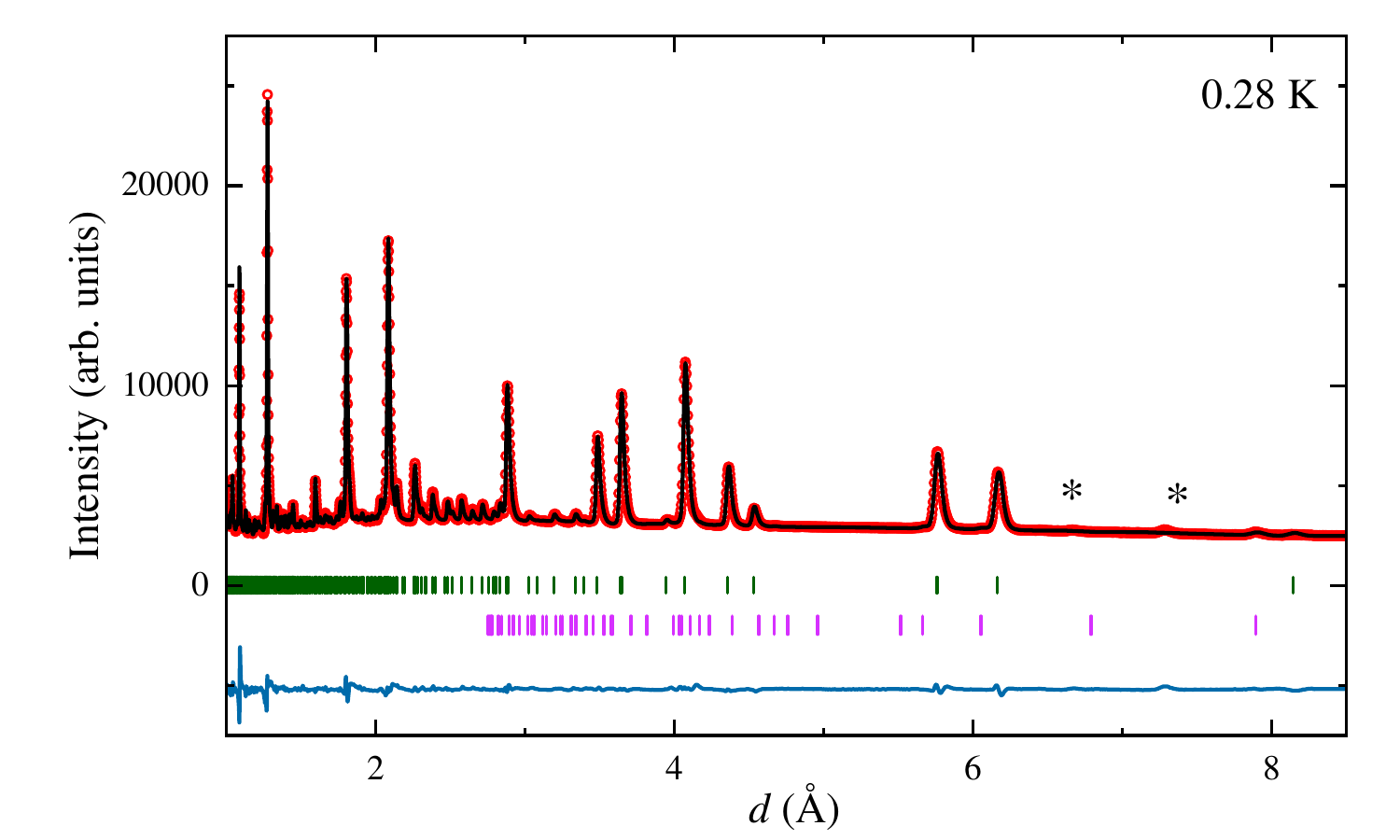}
    \caption{Neutron powder diffraction data (red circles) of the staggered $S=1$ chain Ni(pym)(D$_{2}$O)$_{2}$(NO$_{3}$)$_{2}$ collected at $0.28$\,K. Fits are shown as solid black lines and the solid blue solid line depicts the difference between the observed and fitted intensities. Green and purple tick marks indicate the Bragg peak positions corresponding to the $C2/c$ nuclear phase and magnetic reflections, respectively. The black stars mark low-intensity impurity peaks. The nuclear scattering is 30-40 times more intense than the magnetic scattering, making the magnetic peaks difficult to decern without first subtracting the nuclear contributions, as shown in Fig. 5(b) of the main text.}
    \label{fig: stag_low-T_elastic}
\end{figure}

Elastic neutron diffraction measurements were carried out on the WISH instrument~\cite{Wish} at the ISIS Neutron and Muon Source, UK, using $\sim 1$\,g of partially deuterated powder sample of Ni(pym)(D$_{2}$O)$_{2}$(NO$_{3}$)$_{2}$. Data was collected over a temperature range of $0.28$-$5$\,K, with long counting times of $5.5$\,hours at $0.28$\,K and $5$\,K and 30\,mins collection time was used for intermediate temperatures. Rietveld refinement of the nuclear structure is performed using the FULLPROF software~\cite{Fullprof} using the data collected across four detector banks, at $T=5$\,K. A description of the refinement and structural parameters and comparison to the PXRD parameters is given in Table~\ref{tab: powder_crystal}. The $0.28$\,K data and refinement are presented in Fig~\ref{fig: stag_low-T_elastic}.

\begin{table}
\caption{\label{tab: powder_crystal} Details of the crystal structure refinement of Ni(pym)(DO$_{2}$)$_{3}$(NO$_{3}$)$_{2}$ using data collected on the WISH instrument at $5$\,K.}
\begin{tabular}{p{5cm} p{4.5cm}}
\hline\hline
\textrm{Compound} & Ni(pym)(DO$_{2}$)$_{3}$(NO$_{3}$)$_{2}$\\
\colrule
Emp. formula                    & NiC$_{4}$H$_{4}$D$_{4}$N$_{4}$O$_{8}$ \\
Formula weight\,(g/mol)         & $302.85$     \\
Crystal system, Space group     & Monoclinic, $C2/c$\\
Instrument                      & WISH, 3 detector banks  \\
Method                          & Powder Neutron \\
$T$\,(K)                        & $5$           \\
$a,b,c$\,$(\si{\angstrom})$     & $12.7441(3)$, $11.4933(4)$, $7.387(2)$\\
$\beta$                         & $115.487(3)^{\circ}$\\
$V$\,$(\si{\angstrom}^{3})$     & $976.70(4)$   \\
$Z$                             & $4$            \\
$R_{\text{exp}}$\,($\%$)        & $0.57$         \\
$R_{\text{WP}}$\,($\%$)         & $6.16$         \\
$R_{\text{Bragg}}$\,($\%$)      & $4.952$       \\
\hline\hline
\end{tabular}
\end{table}

\subsection{Inelastic neutron scattering}

Inelastic neutron scattering measurements were performed on the LET instrument~\cite{LET} at the ISIS Neutron and Muon Source, UK, using $\sim 1$\,g of partially deuterated powder sample of Ni(pym)(D$_{2}$O)$_{2}$(NO$_{3}$)$_{2}$. Data was collected at $0.26$\,K, $4.31$\,K, $21.11$\,K and $50$\,K, with exposure times of $16$ hours, $13$ hours, $5$ hours and $6$ hours respectively, using incident energies $e_{i} = 2.04$\,meV, $3.36$\,meV, $6.55$\,meV and $18$\,meV. 

\begin{figure}[t]
    \centering
    \includegraphics[width= 0.75\textwidth]{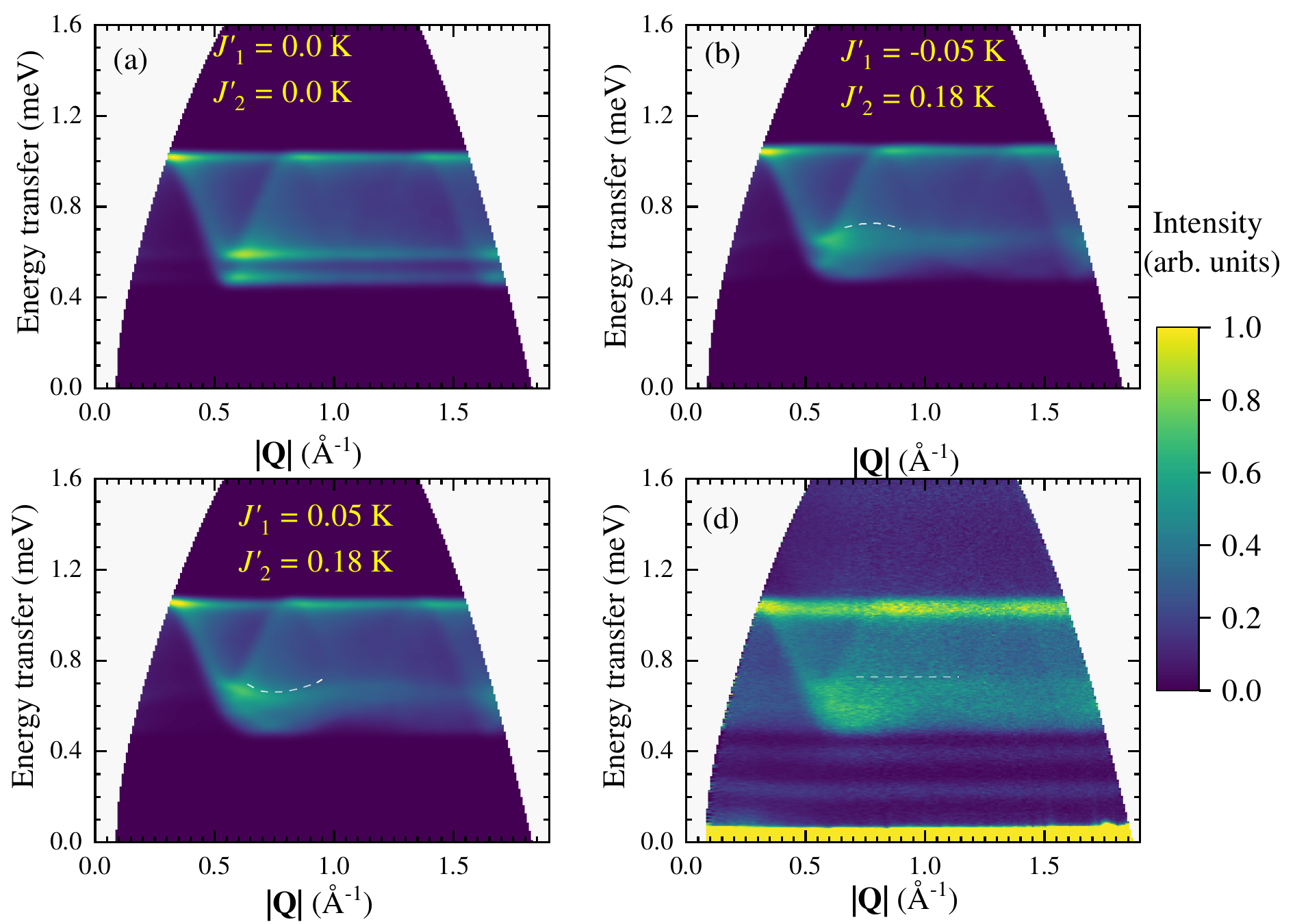}
\caption{ Linear spin wave theory simulations of the spin-wave spectra for Ni(pym)(D$_{2}$O)$_{2}$(NO$_{3}$)$_{2}$ using Eq.~1 and different values of $J'_{1}$ and $J'_{2}$. The simulation with Hamiltonian parameters $J_{0}=5.107$\,K,  $D = -2.79$\,K, $E=0.19$\,K and interchain interaction: (a) $J'_{1}=0$\,K and $J'_{2}=0$\,K, (b) $J'_{1}=-0.05$\,K (antiferromagnetic) and $J'_{2}=0.18$\,K and (c) $J'_{1}=0.05$\,K (ferromagnetic) and $J'_{2}=0.18$\,K. (d) Data collected at $T=0.26$\,K and with $\epsilon_{i}=2.04$\,meV. The white dashed line shows the dispersion of spin-wave modes at $0.76$\,meV, due to a non-zero $J'_{1}$, which appears flat in the data.}
\label{fig: Stag_let_maps2}
\end{figure}

Figure~\ref{fig: Stag_let_maps2}(a)-(c) present simulations of the spin-wave spectra using different values of the interchain interaction $J'_{1}$ and $J'_{2}$. When $J'_{1}=0$\,K and $J'_{2}=0$\,K, the modulation of the gap as a function of $|\mathbf{Q}|$ is not present and dispersionless interchain excitations are found. Adding the $J'_{1}$ term results in the dispersion of the the band at $0.76$\,meV, as shown by the white dashed lines in Fig.~\ref{fig: Stag_let_maps2}(b) and (c). This band appears to be flat in the data [see Fig.~\ref{fig: Stag_let_maps2}(d)] and suggests that $J'_{1}$ is small.

\section{Calculations}
\subsection{Anisotropy tensor}

We use a site-dependent local anisotropy tensor, $K_{i}$ in the global $xyz$ laboratory frame to account for the staggered octahedral orientation. For LSWT simulation in SpinW~\cite{SpinW}, this laboratory frame is chosen such that $x$ is parallel to the crystallographic $a$-axis and $y$ is parallel to the $b$-axis. The anisotropy energy of Ni(II) ions in the frame of the local octahedra is 
\begin{equation}
    \mathcal{H}_{\text{SIA}} = \mathbf{S}^{T}K^{\text{loc}}\mathbf{S} =
        \begin{pmatrix}
        S_{x}&S_{y}&S_{z}
    \end{pmatrix}
    \begin{bmatrix}
        E& 0& 0\\
        0&-E& 0\\
        0& 0& D
    \end{bmatrix}
    \begin{pmatrix}
        S_{x}\\S_{y}\\S_{z}
    \end{pmatrix}.
\end{equation}
The local $K^{\text{loc}}$ tensors are transformed to the global frame using Euler rotations
\begin{equation}
    K_{i} = Q_{i}^{T}K^{\text{loc}}Q_{i},
\end{equation}
where
\begin{equation}
    Q_{i} = \begin{bmatrix}
        \cos(\gamma)&(-1)^{i}\sin(\gamma)& 0\\
        (-1)^{i+1}\sin(\gamma)& \cos(\gamma)& 0\\
        0           & 0           & 1
    \end{bmatrix}
    \begin{bmatrix}
        1         & 0              & 0\\
        0         & \cos(\alpha)  & (-1)^{i}\sin(\alpha)\\
        0         &(-1)^{i+1}\sin(\alpha)  & \cos(\alpha)
    \end{bmatrix}
    \begin{bmatrix}
        \cos(\rho)& 0           & (-1)^{i+1}\sin(\rho)\\
        0         & 1           & 0\\
       (-1)^{i}\sin(\rho)& 0           & \cos(\rho)
    \end{bmatrix}
\end{equation}

The angles $\alpha=33.88(4)^{\circ}$, $\rho=58.00(1)^{\circ}$ measures the tilting of the Ni-N axials bonds away from the $x$ and $z$ directions respectively and $\gamma = 41.81(4)^{\circ}$ is the rotation of the octahedra about the axial bond. These angles are determined from the single-crystal XRD structure.

\subsection{Magnetic structure factor}

{\renewcommand{\arraystretch}{1.2}
\begin{table}[h]
\centering
\caption{\label{tab: stag_sites} Positions ($r_{i}$) of Ni(II) spin sites within the $C2/c$ crystallographic unit cell and the decomposition of the magnetic moments ($m_{i}$) into ferromagnetic ($m_{\text{FM}}$) and antiferromagnetic ($m_{\text{AFM}}$) modes described by the symmetry of $mL^{+}_{1}$ irrep.}
\begin{tabular}{m{2cm} m{4cm} m{4.5cm}}
\hline\hline
\textrm{Ni(II) site}&
\textrm{position $(x,y,z)$}&
\textrm{magnetic moment}\\
\colrule
 1 & $\textbf{r}_{1} = \left[\frac{1}{4}\,\frac{3}{4}\,0\right]$ & $\textbf{m}_{1} = +\textbf{m}_{FM}+\textbf{m}_{AFM}$\\
 2 & $\textbf{r}_{2} = \left[\frac{1}{4}\,\frac{1}{4}\,\frac{1}{2}\right]$ & $\textbf{m}_{2} = -\textbf{m}_{FM}+\textbf{m}_{AFM}$\\
 3 & $\textbf{r}_{3} = \left[\frac{3}{4}\,\frac{1}{2}\,0\right]$ & $\textbf{m}_{3} = \textbf{m}_{1} = +\textbf{m}_{FM}+\textbf{m}_{AFM}$\\
 4 & $\textbf{r}_{4} = \left[\frac{3}{4}\,\frac{3}{4}\,\frac{1}{2}\right]$ & $\textbf{m}_{4} = -\textbf{m}_{2} = +\textbf{m}_{FM}-\textbf{m}_{AFM}$\\
\hline\hline
\end{tabular}
\end{table}}

Within the crystallographic $C/2c$ unit cell of the Ni(pym)(H$_{2}$O)$_{2}$(NO$_{3}$)$_{2}$ nuclear structure, there are four Ni(II) spins residing at the given in Table.~\ref{tab: stag_sites}. The magnetic structure factor for a magnetic satellite peaks at $\mathbf{Q_{k}} = (h,k,l)+\mathbf{k}$ is
\begin{equation}
    \mathbf{F}(\mathbf{Q_{k}}) \propto \sum_{i=1}^{4}\textbf{m}_{i}\exp{2\pi i(\mathbf{Q_{k}}\cdot\mathbf{r_{i}})},
    \label{eq: struct_fact}
\end{equation}
where $\textbf{m}_{i}$ were magnetic moments of the site at unit cell position $\mathbf{r_{i}}$. Sites related by $c$-centering ($[+1/2,\,+1/2,\,0]$) are constrained to point antiparallel, while sites connected by pym are not symmetrically constrained in the $mL^{+}_{1}$ irrep. Applying these symmetry constraints and the observed $h+k=2n$ reflection condition gives $\mathbf{F}(\mathbf{Q_{k}})$, in terms of the two sites connected by pym ligands, 
\begin{equation}
    \mathbf{F}(\mathbf{Q_{k}}) \propto \textbf{m}_{1}\exp{2\pi i(\mathbf{Q_{k}}\cdot\mathbf{r_{1}})} + \textbf{m}_{4}\exp{2\pi i(\mathbf{Q_{k}}\cdot\mathbf{r_{4}})}
\end{equation}
\begin{equation}
    \mathbf{F}(\mathbf{Q_{k}}) \propto \exp{2\pi i(\mathbf{Q_{k}}\cdot\mathbf{r_{1}})} \left [ \textbf{m}_{1} + \textbf{m}_{4}\exp{2\pi i(\mathbf{Q_{k}}\cdot\left[\mathbf{r_{4}}-\mathbf{r_{1}}\right]}\right ]
\end{equation}
\begin{equation}
    \mathbf{F}(\mathbf{Q_{k}}) \propto \exp{2\pi i(\mathbf{Q_{k}}\cdot\mathbf{r_{1}})} \left [ \textbf{m}_{1} - \textbf{m}_{4}\exp{\pi i\left[h+l\right]}\right].
    \label{eq: struct_fact}
\end{equation}
In magnetic neutron scattering experiments, the scattering intensity is proportional to the projection of $\mathbf{F}(\mathbf{Q_{k}})$ into the plane perpendicular to the $\mathbf{Q_{k}}$ and is given by
\begin{equation}
    I \propto |\mathbf{F}_{\perp}|^{2} = \mathbf{F}^{*}\cdot\mathbf{F} - (\mathbf{F}^{*}\cdot\mathbf{\hat{Q}_{k}})(\mathbf{F}\cdot\mathbf{\hat{Q}_{k}}),
    \label{eq: intensity}
\end{equation}
where $\mathbf{\hat{Q}_{k}}$ is the scattering unit vector~\cite{boothroyd2020principles}. The moments of these two spins can be decomposed into a linear combination of real orthogonal FM ($\textbf{m}_{\text{FM}}$) and AFM ($\textbf{m}_{\text{AFM}}$) modes: 
\begin{equation}
\begin{split}
    \textbf{m}_{1} &= \textbf{m}_{\text{FM}} + \textbf{m}_{\text{AFM}}, \\
    \textbf{m}_{4} &= \textbf{m}_{\text{FM}} - \textbf{m}_{\text{AFM}},
\end{split}
\label{eq: model_decopm}
\end{equation}
Substituting Eq.~\ref{eq: struct_fact} and ~\ref{eq: model_decopm} into Eq.~\ref{eq: intensity} gives
\begin{equation}
\begin{split}
    \mathbf{F}^{*}\cdot\mathbf{F} \propto &
    (\textbf{m}_{\text{FM}}[1 - \exp{-\pi i\left[h+l\right]}] +\textbf{m}_{\text{AFM}}[1 + \exp{-\pi i\left[h+l\right]}])\cdot\\
    &(\textbf{m}_{\text{FM}}[1 - \exp{\pi i\left[h+l\right]}] +\textbf{m}_{\text{AFM}}[1 + \exp{\pi i\left[h+l\right]}]),
\end{split}
\end{equation}
and
\begin{equation}
\begin{split}
    (\mathbf{F}^{*}\cdot\mathbf{\hat{Q}_{k}})(\mathbf{F}\cdot\mathbf{\hat{Q}_{k}}) \propto &(\textbf{m}_{\text{FM}}\cdot\mathbf{\hat{Q}_{k}}[1 - \exp{\pi i\left[h+l\right]}] +\textbf{m}_{\text{AFM}}\cdot\mathbf{\hat{Q}_{k}}[1 + \exp{\pi i\left[h+l\right]}])\cdot\\
    &(\textbf{m}_{\text{FM}}\cdot\mathbf{\hat{Q}_{k}}[1 - \exp{-\pi i\left[h+l\right]}]+\textbf{m}_{\text{AFM}}\cdot\mathbf{\hat{Q}_{k}}[1 + \exp{-\pi i\left[h+l\right]}]).
\end{split}
\end{equation}
Simplifying yields,
\begin{equation}
\begin{split}
        I \propto &[|\textbf{m}_{\text{FM}}|^{2} - |\textbf{m}_{\text{FM}}\cdot\mathbf{\hat{Q}_{k}}|^{2}]\cdot[2-2\cos(\pi(h+l))] +\\
        &[|\textbf{m}_{\text{AFM}}|^{2} - |\textbf{m}_{\text{AFM}}\cdot\mathbf{\hat{Q}_{k}}|^{2}]\cdot[2+2\cos(\pi(h+l))].
\end{split}
\label{eq: intensity_2}
\end{equation} 
From Eq.~\ref{eq: intensity_2} it can be seen that for $h+l = 2n-1$ the FM mode contribution to the intensity is zero and for $h+l = 2n$ the AFM mode contribution is zero.
\subsection{Canting angle estimate}
\begin{figure}[h]
    \centering
    \includegraphics[width= 0.75\linewidth]{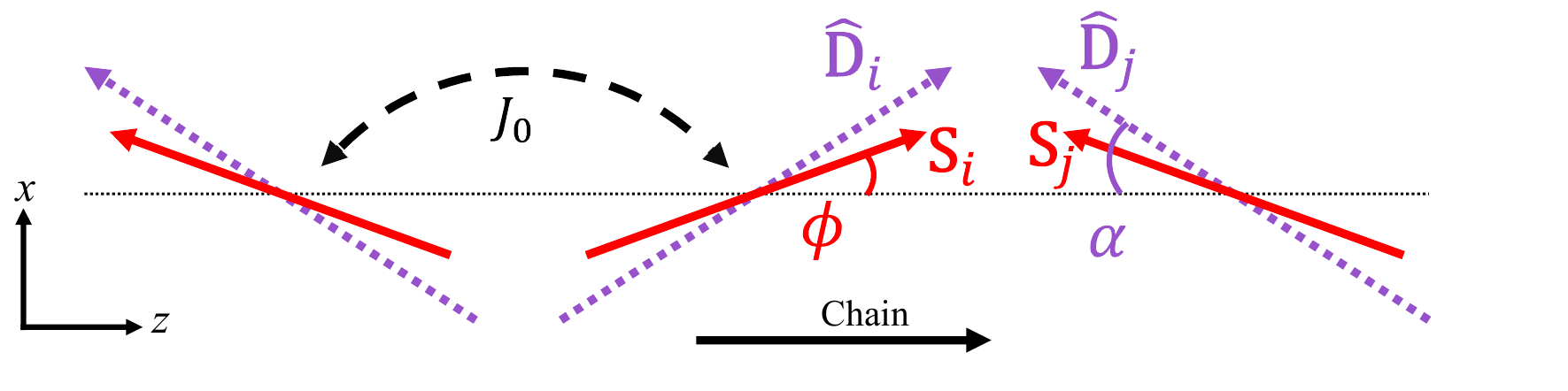}
    \caption{Schematic representation of the Ni staggered chain Ni(pym)(D$_{2}$O)$_{2}$(NO$_{3}$)$_{2}$. The red arrows represent the magnetic moment direction with canting angle $\phi$ and the purple arrows show the local single-ion-anisotropy direction at angle $\alpha$ from the chain direction.}
    \label{fig: min_Hamitonian}
\end{figure}

Similar to Ref.~\cite{Pianet_2017_DW}, we employ a minimal Hamiltonian containing only $J_{0}$ and $D$ terms,
\begin{equation}
   \mathcal{H}=J_{0}\sum_{\left \langle i,j \right \rangle}\mathbf{S}_{i}\cdot\mathbf{S}_{j} +D \left[\sum_{i}\left (\hat{\mathbf{D}}_{i}\cdot\mathbf{S}_{i} \right )^{2}
     +\sum_{j} \left (\hat{\mathbf{D}}_{j}\cdot\mathbf{S}_{j} \right )^{2} \right],
    \label{eq: min_Hamitonian}
\end{equation}
to estimate the SIA-induced canting angle ($\phi$) for Ni(pym)(H$_{2}$O)$_{2}$(NO$_{3}$)$_{2}$. Here $\mathbf{S}_{i,j}$ is the spin vectors of each ion, $\left \langle i,j \right \rangle$ denotes the sum over unique newerest neighbour exchange bonds, $\hat{\mathbf{D}_{i}} = [\sin(\alpha), \cos(\alpha)]$ and $\hat{\mathbf{D}}_{j} = [-\sin(\alpha), \cos(\alpha) ]$ were two dimensional SIA unit vectors for sites $i$ and $j$. A schematic of the configuration described by this Hamiltonian is shown in Fig.~\ref{fig: min_Hamitonian}. The ground state energy is
\begin{equation}
    \epsilon = -\frac{Jn}{2}\cos(2\phi) + D\cos^2(\phi-\alpha).
\end{equation}
Here, $n=2$ is the number of neighbouring ions and the factor of $1/2$ accounts for the double counting of the exchange bonds. Minimising this energy with respect to $\phi$ yields
\begin{equation}
    \frac{D}{J_{0}} = \frac{\sin{2\phi}}{\sin(\phi-\alpha)\cos(\phi-\alpha)}.
    \label{supp_eq: DJ_ratio}
\end{equation}

\subsection{Monte-Carlo simulations}

Monte-Carlo simulations of $M(H)$ were performed using a routine written in Julia~\cite{bezanson2017julia}. The simulation utilises a cluster of 8 ions arranged in the configuration shown in Fig.~6(a) in the main text, with periodic boundary conditions. The magnetic moment of each ion is represented as a classical unit vector and the simulations aim to minimise the ground-state energy of the Hamiltonian in Eq.~1 at each value of $\mu_{0}H$ using simulated annealing loops. The powder average $M(H)$ is calculated by averaging the magnetisation curves for all 100 different field orientations, which are evenly spaced on a unit sphere using the golden ratio method~\cite{Hannay_2004_fibo}.

\bibliography{main.bib}